\documentclass[
  journal=pasa,
  manuscript=research-paper, 
  year=202X,
  volume=YY,
]{cup-journal}

\usepackage{amsmath}
\usepackage{amssymb,microtype,siunitx,booktabs}

\usepackage{comment}
\usepackage{upgreek}
\usepackage{ulem}
\usepackage{hyperref}

\title[Understanding the CE drag torque]{Understanding the Drag Torque in Common Envelope Evolution}

\author{Soumik Bhattacharyya}
\affiliation{National Institute of Science Education and Research, An OCC of Homi Bhabha National Institute, Bhubaneswar 752050, Odisha, India}

\author{Luke Chamandy}
\affiliation{National Institute of Science Education and Research, An OCC of Homi Bhabha National Institute, Bhubaneswar 752050, Odisha, India}
\alsoaffiliation{Department of Physics and Astronomy, University of Rochester, Rochester, NY 14627, USA}

\author{Eric G. Blackman}
\affiliation{Department of Physics and Astronomy, University of Rochester, Rochester, NY 14627, USA}
\email[E. G. Blackman]{blackman@pas.rochester.edu}

\author{Adam Frank}
\affiliation{Department of Physics and Astronomy, University of Rochester, Rochester, NY 14627, USA}

\author{Baowei Liu}
\affiliation{Department of Physics and Astronomy, University of Rochester, Rochester, NY 14627, USA}
\alsoaffiliation{Center for Integrated Research Computing, University of Rochester, Rochester, NY 14627, USA}

\received {dd Mmm YYYY}
\revised  {dd Mmm YYYY}
\accepted {dd Mmm YYYY}
\published{XX XXX 202X}


\defcitealias{Escala+04}{E04}
\defcitealias{Kim+08}{K08}

\begin{document}

\def\apj{ApJ}
\def\apss{Ap{\&}SS}
\def\mnras{MNRAS}
\def\aap{A\&A}
\def\apjl{ApJ}
\def\gafd{GAFD}
\def\jfm{JFM}
\def\physrep{PhR}
\def\pre{PhRvE}
\def\prl{PhRvL}
\def\apjs{ApJS}
\def\pasa{PASA}
\def\pasj{PASJ}
\def\nat{Nature}
\def\pasp{PASP}
\def\ssr{SSRv}
\def\araa{ARA\&A}
\def\aj{AJ}
\def\sgg{Stud.\ Geophys.\ Geod.}
\def\na{New\ Astron.}
\def\aapr{ARA\&A}
\def\gapfd{GApFD}
\def\an{AN}
\renewcommand{\vec}[1]{{{\mbox{\boldmath $#1$}}}}
\newcommand{\ar}{_\mathrm{a}}
\newcommand{\aar}{\mathrm{,a}}
\newcommand{\ir}{_\mathrm{i}}
\newcommand{\iir}{\mathrm{,i}}
\newcommand{\f}{_\textrm{0}}
\newcommand{\sign}[1]{\textrm{sign}({#1})}
\newcommand{\zhat}{\hat{z}}
\newcommand{\bfxhat}{\vec{\hat{x}}}
\newcommand{\bfyhat}{\vec{\hat{y}}}
\newcommand{\bfzhat}{\vec{\hat{z}}}
\newcommand{\bfRhat}{\vec{\hat{R}}}
\newcommand{\bfphihat}{\vec{\hat{\phi}}}
\newcommand{\What}{\widehat{W}}
\newcommand{\Vhat}{\widehat{V}}
\newcommand{\Exp}[1]{{\rm e}^{#1}}
\newcommand{\Log}{\mathrm{Log}}
\newcommand{\del}{\partial}
\newcommand{\Del}{{\nabla}}
\newcommand{\bfDel}{\vec{\nabla}}
\newcommand{\bmDel}{\vec{\nabla}}
\newcommand{\bfomega}{\vec{\omega}}
\newcommand{\bmomega}{\vec{\omega}}
\newcommand{\bfOmega}{\vec{\Omega}}
\newcommand{\bmOmega}{\vec{\Omega}}
\newcommand{\bfgamma}{\vec{\gamma}}
\newcommand{\bmgamma}{\vec{\gamma}}
\newcommand{\alp}{\alpha}
\newcommand{\Gam}{\Gamma}
\newcommand{\eps}{\epsilon}
\newcommand{\Delsq}{\nabla^2}
\newcommand{\Emf}{\vec{\mathcal{E}}}
\newcommand{\Fmf}{\vec{F}}
\newcommand{\Flux}{\vec{\mathcal{F}}}
\newcommand{\flux}{\mathcal{F}}
\newcommand{\Bs}{\mathcal{B}}
\newcommand{\Bsr}{\mathcal{B}_r}
\newcommand{\Bsp}{\mathcal{B}_\phi}
\newcommand{\Bsz}{\mathcal{B}_z}
\newcommand{\bfU}{\vec{U}}
\newcommand{\bfB}{\vec{B}}
\newcommand{\bfJ}{\vec{J}}
\newcommand{\bfu}{\vec{u}}
\newcommand{\bfv}{\vec{v}}
\newcommand{\bfb}{\vec{b}}
\newcommand{\bmb}{\vec{b}}
\newcommand{\bfj}{\vec{j}}
\newcommand{\bmj}{\vec{j}}
\newcommand{\bfr}{\vec{r}}
\newcommand{\bfp}{\vec{\phi}}
\newcommand{\bfz}{\vec{z}}
\newcommand{\bfx}{\vec{x}}
\newcommand{\bfF}{\vec{F}}
\newcommand{\bfs}{\vec{s}}
\newcommand{\Emfr}{\mathcal{E}_r}
\newcommand{\Emfp}{\mathcal{E}_\phi}
\newcommand{\Emfz}{\mathcal{E}_z}
\newcommand{\Emfi}{\mathcal{E}_i}
\newcommand{\Fmfr}{F_r}
\newcommand{\Fmfp}{F_\phi}
\newcommand{\Fmfz}{F_z}
\newcommand{\Fluxr}{\mathcal{F}_r}
\newcommand{\Fluxphi}{\mathcal{F}_\phi}
\newcommand{\Fluxz}{\mathcal{F}_z}
\newcommand{\mean}[1]{\overline{#1}}
\newcommand{\meanv}[1]{\vec{#1}}
\newcommand{\avg}[1]{\left<{#1}\right>}
\newcommand{\corr}{_\mathrm{c}}						
\newcommand{\corot}{_\mathrm{c}}						
\newcommand{\D}{_\mathrm{D}}						
\newcommand{\eq}{_\mathrm{eq}}						
\newcommand{\eqdisk}{_\mathrm{eq,d}}						
\newcommand{\eqhalo}{_\mathrm{eq,h}}						
\newcommand{\rms}{_\mathrm{0}}						
\newcommand{\forc}{_\mathrm{f}}					   	
\newcommand{\fdisk}{_\mathrm{0,d}}					   	
\newcommand{\fhalo}{_\mathrm{0,h}}					   	
\newcommand{\z}{_\mathrm{0}}					   	
\newcommand{\1}{_\mathrm{1}}					   	
\newcommand{\core}{_\mathrm{1,c}}					   	
\newcommand{\env}{_\mathrm{env}}					   	
\newcommand{\rot}{_\mathrm{rot}}					   	
\newcommand{\2}{_\mathrm{2}}					   	
\newcommand{\kin}{_\mathrm{k}}			   		
\newcommand{\magn}{_\mathrm{m}}			   		
\newcommand{\turb}{_\mathrm{t}}			   		
\newcommand{\etat}{\eta_\mathrm{t}}			   	
\newcommand{\kappat}{\kappa_\mathrm{t}}			   	
\newcommand{\Turb}{\mathrm{t}}			   		
\newcommand{\crit}{_\mathrm{c}}			   		
\newcommand{\sat}{_\mathrm{sat}}			   		
\newcommand{\magnsat}{_\mathrm{m,sat}}			   		
\newcommand{\const}{\mathrm{const}}			   	
\newcommand{\mx}{\mathrm{max}}			   		
\newcommand{\ma}{_\mathrm{max}}			   		
\newcommand{\dd}{\mathrm{d}}			   		
\newcommand{\diff}{_\mathrm{d}}			   		
\newcommand{\udiff}{^\mathrm{d}}			   		
\newcommand{\pol}{_\mathrm{p}}			   		
\newcommand{\diffzero}{_\mathrm{d,0}}
\newcommand{\Adv}{\mathrm{a}}			   		
\newcommand{\Diff}{\mathrm{d}}			   		
\newcommand{\VC}{_\mathrm{VC}}			   		
\newcommand{\uVC}{^\mathrm{VC}}			   		
\newcommand{\on}{_\mathrm{on}}
\newcommand{\off}{_\mathrm{off}}
\newcommand{\cro}{\times}
\newcommand{\Rm}{\mathcal{R}_\mathrm{m}}
\newcommand{\Rey}{\mathcal{R}_\mathrm{e}}
\newcommand{\mbr}{B_r}
\newcommand{\mbp}{B_\phi}
\newcommand{\mbz}{B_z}
\newcommand{\mbi}{B_i}
\newcommand{\mur}{U_r}
\newcommand{\mup}{U_\phi}
\newcommand{\muz}{U_z}
\newcommand{\muP}{U_\mathrm{p}}
\newcommand{\muztilde}{\widetilde{\mean{U}}_z}
\newcommand{\mui}{\mean{U}_i}
\newcommand{\tautilde}{\widetilde{\tau}}
\newcommand{\gammatilde}{\widetilde{\gamma}}
\newcommand{\alphatilde}{\widetilde{\alpha}}
\newcommand{\betatilde}{\widetilde{\beta}}
\newcommand{\kappatilde}{\widetilde{\kappa}}
\newcommand{\alptilde}{\alphatilde}
\newcommand{\lambdatilde}{\widetilde{\lambda}}
\newcommand{\lamtilde}{\lambdatilde}
\newcommand{\omegatilde}{\widetilde{\omega}}
\newcommand{\omtilde}{\omegatilde}
\newcommand{\Ctilde}{\widetilde{C}}
\newcommand{\rtilde}{\widetilde{r}}
\newcommand{\qtilde}{\widetilde{q}}
\newcommand{\atilde}{\widetilde{a}}
\newcommand{\btilde}{\widetilde{b}}
\newcommand{\Atilde}{\widetilde{A}}
\newcommand{\Btilde}{\widetilde{B}}
\newcommand{\mbrtilde}{\widetilde{B}_r}
\newcommand{\mbptilde}{\widetilde{B}_\phi}
\newcommand{\Dtilde}{\widetilde{D}}
\newcommand{\xitilde}{\widetilde{\xi}}
\newcommand{\etatilde}{\widetilde{\eta}}
\newcommand{\Vpot}{\mathcal{V}}
\renewcommand{\theenumi}{\arabic{enumi})}
\renewcommand{\labelenumi}{\theenumi}
\newcommand{\alphabar}{\mean{\alpha}}
\newcommand{\rci}{r_{\mathrm{c},i}}
\newcommand{\I}{_\mathrm{I}}
\newcommand{\pat}{_\mathrm{p}}
\newcommand{\arm}{_\mathrm{a}}
\newcommand{\interarm}{_\mathrm{i}}
\newcommand{\critarm}{_\mathrm{c,a}}
\newcommand{\critinterarm}{_\mathrm{c,i}}
\newcommand{\uarm}{_\mathrm{U,a}}
\newcommand{\uinterarm}{_\mathrm{U,i}}
\newcommand{\kappaarm}{_\mathrm{\kappa,a}}
\newcommand{\kappainterarm}{_\mathrm{\kappa,i}}
\newcommand{\disk}{_\mathrm{d}}
\newcommand{\disc}{_\mathrm{d}}
\newcommand{\halo}{_\mathrm{h}}
\newcommand{\Sr}{S}
\newcommand{\Sz}{S_{z}}
\newcommand{\Coriolis}{\mathrm{Co}}
\newcommand{\Strouhal}{\mathrm{St}}
\newcommand{\Ma}{\mathcal{M}}
\newcommand{\Pm}{\mathrm{Pm}}
\newcommand{\cyc}{_\mathrm{cyc}}
\newcommand{\cut}{_\mathrm{c}}
\newcommand{\mesa}{_\mathrm{\textsc{MESA}}}
\newcommand{\gas}{_\mathrm{gas}}
\newcommand{\dynam}{_\mathrm{dyn}}
\newcommand{\oneenv}{_\mathrm{1,e}}
\newcommand{\twoenv}{_\mathrm{2,e}}
\newcommand{\soft}{_\mathrm{soft}}
\newcommand{\softf}{_\mathrm{soft,0}}
\newcommand{\interior}{_\mathrm{in}}
\newcommand{\inone}{_\mathrm{in,1}}
\newcommand{\intwo}{_\mathrm{in,2}}
\newcommand{\Kep}{_\mathrm{K}}
\newcommand{\out}{_\mathrm{out}}
\newcommand{\surf}{_\mathrm{surf}}
\newcommand{\sink}{_\mathrm{sink}}
\newcommand{\launch}{_\mathrm{launch}}
\newcommand{\snap}{_\mathrm{snap}}
\newcommand{\jet}{_\mathrm{jet}}
\newcommand{\Gn}{\mathrm{G}}
\newcommand{\init}{_\mathrm{i}}
\newcommand{\final}{_\mathrm{f}}
\newcommand{\acc}{_\mathrm{acc}}
\newcommand{\Edd}{_\mathrm{Edd}}
\newcommand{\mass}{_\mathrm{m}}
\newcommand{\angmom}{_\mathrm{a}}
\newcommand{\proton}{_\mathrm{p}}
\newcommand{\Rsol}{\,\mathrm{R_\odot}}
\newcommand{\Msol}{\,\mathrm{M_\odot}}
\newcommand{\Rsun}{\,\mathrm{R_\odot}}
\newcommand{\Msun}{\,\mathrm{M_\odot}}
\newcommand{\Msunyr}{\,\mathrm{M_\odot\,yr^{-1}}}
\newcommand{\base}{_\mathrm{base}}
\newcommand{\amb}{_\mathrm{amb}}
\newcommand{\bo}{_\mathrm{box}}
\newcommand{\starmax}{_{*,\mathrm{max}}}
\newcommand{\rmD}{\mathrm{\Delta}}
\newcommand{\rmdelta}{\mathrm{\delta}}
\newcommand{\CE}{_\mathrm{CE}}
\newcommand{\bx}{_\mathrm{box}}
\newcommand{\bind}{_\mathrm{bind}}
\newcommand{\orb}{_\mathrm{orb}}
\newcommand{\refine}{_\mathrm{ref}}
\newcommand{\potential}{_\mathrm{pot}}
\newcommand{\kinetic}{_\mathrm{kinetic}}
\newcommand{\internal}{_\mathrm{internal}}
\newcommand{\rmd}{\mathrm{d}}
\newcommand{\tot}{_\mathrm{tot}}
\newcommand{\unb}{_\mathrm{unb}}
\newcommand{\conv}{_\mathrm{conv}}
\newcommand{\adv}{_\mathrm{adv}}
\newcommand{\sound}{_\mathrm{s}}
\newcommand{\Kepler}{_\mathrm{K}}
\newcommand{\abar}{\overline{a}}
\newcommand{\kB}{k_\mathrm{B}}
\newcommand{\mHe}{m_\mathrm{He}}
\newcommand{\rec}{_\mathrm{rec}}
\newcommand{\thm}{_\mathrm{thm}}
\newcommand{\HeI}{\mathrm{He\,I}}
\newcommand{\HeII}{\mathrm{He\,II}}
\newcommand{\HeIII}{\mathrm{He\,III}}
\newcommand{\HI}{\mathrm{H\,I}}
\newcommand{\HII}{\mathrm{H\,II}}
\newcommand{\Hmol}{\mathrm{H_2}}
\newcommand{\bnd}{_\mathrm{bnd}}
\newcommand{\rad}{_\mathrm{rad}}
\newcommand{\rhobar}{\overline{\rho}}
\newcommand{\pert}{_\mathrm{p}}
\newcommand{\Mach}{\mathcal{M}}
\newcommand{\CM}{_\mathrm{CM}}
\newcommand{\ff}{_\mathrm{ff}}

\newcommand\bgreek[1]{ \mathchoice
    {\hbox{\boldmath$\displaystyle{#1}$\unboldmath}}%
    {\hbox{\boldmath$\textstyle{#1}$\unboldmath}}%
    {\hbox{\boldmath$\scriptstyle{#1}$\unboldmath}}%
    {\hbox{\boldmath$\scriptscriptstyle{#1}$\unboldmath}}}
\newcommand{\lukec}[1]{\textcolor{purple}{[Luke comments: #1]}}
\newcommand{\luke}[1]{\textcolor{red}{#1}}
\newcommand{\eric}[1]{\textcolor{brown}{[Eric: #1]}}
\newcommand{\adam}[1]{\textcolor{green!50!blue}{[Adam: #1]}}
\newcommand{\bliuc}[1]{\textcolor{teal}{[Baowei: #1]}}
\newcommand{\bliu}[1]{\textcolor{purple}{#1}}
\newcommand{\soumikc}[1]{\textcolor{olive}{[\textbf{Soumik comments:} #1]}}
\newcommand{\soumik}[1]{\textcolor{red}{#1}}

%
%
  \renewcommand{\cm}{\,{\rm cm}}
  \newcommand{\cmcmcm}{\,{\rm cm^{-3}}}
  \newcommand{\dyn}{\,{\rm dyn}}
  \newcommand{\dyncmcmcm}{\,{\rm dyn\,cm^{-3}}}
  \newcommand{\erg}{\,{\rm erg}}
  \renewcommand{\g}{\,{\rm g}}
  \newcommand{\gcmcmcm}{\,{\rm g\,cm^{-3}}}
  \newcommand{\Jy}{\,{\rm Jy}}
  \newcommand{\Jyb}{\,{\rm Jy/beam}}
  \renewcommand{\km}{\,{\rm km}}
  \newcommand{\kms}{\,{\rm km\,s^{-1}}}
  \newcommand{\kmskpc}{\,{\rm km\,s^{-1}\,kpc^{-1}}}
  \newcommand{\kmskpckpc}{\,{\rm km\,s^{-1}\,kpc^{-2}}}
  \newcommand{\kmskpckpckpc}{\,{\rm km\,s^{-1}\,kpc^{-3}}}
  \newcommand{\cmcms}{\,{\rm cm^2\,s^{-1}}}
  \newcommand{\mJy}{\,{\rm mJy}}
  \newcommand{\mJyb}{\,{\rm mJy/beam}}
  \newcommand{\kpc}{\,{\rm kpc}}
  \newcommand{\pc}{\,{\rm pc}}
  \newcommand{\Mpc}{\,{\rm Mpc}}
  \newcommand{\Myr}{\,{\rm Myr}}
  \newcommand{\iMyr}{\,{\rm Myr^{-1}}}
  \newcommand{\Gyr}{\,{\rm Gyr}}
  \newcommand{\iGyr}{\,{\rm Gyr^{-1}}}
  \newcommand{\mG}{\,{\rm mG}}
  \newcommand{\mkG}{\,\mu{\rm G}}
  \newcommand{\nG}{\,{\rm nG}}
  \newcommand{\p}{\,{\rm pc}}
  \newcommand{\radm}{\,{\rm rad\,m^{-2}}}
  \renewcommand{\s}{\,{\rm s}}
  \newcommand{\yr}{\,{\rm yr}}     
  \newcommand{\da}{\,{\rm d}}     
  \newcommand{\dyncmcm}{\,{\rm dyn\,cm^{-2}}}     
  \newcommand{\dynecmcm}{\,{\rm dyn\,cm^{-2}}}     
  \newcommand{\au}{\,{\rm AU}}     
  \newcommand{\amu}{\,\mathrm{amu}}
  \newcommand{\ergd}{\,{\rm erg\,d^{-1}}}
  

\begin{abstract}
Common envelope (CE) evolution is largely governed by the drag torque applied on the in-spiralling stellar components by the envelope. 
Previous work has shown that idealized models of the torque based on a single body moving in rectilinear motion 
through an unperturbed atmosphere can be highly inaccurate. 
Progress requires new models for the torque that account for binarity. 
Toward this end we perform a new 3D global hydrodynamic CE simulation with the mass of the companion point particle 
set equal to the mass of the asymptotic giant branch star core particle to maximize symmetry and facilitate interpretation. 
First, 
we find that a region around the particles of a scale comparable to their separation contributes essentially all of the torque. 
Second, the density pattern of the torque-dominating gas and, to an extent, this
gas itself, is roughly in corotation with the binary. 
Third, approximating the spatial distribution of the torquing gas as a uniform-density prolate spheroid whose major axis resides in the orbital plane and lags the line joining the binary components by a constant phase angle reproduces the torque evolution remarkably well, analogous to studies of binary supermassive black holes. 
Fourth, we compare the torque measured in the simulation with the predictions of a model 
that assumes two weak point-mass perturbers undergoing circular motion in a uniform background
without gas self-gravity, and find remarkable agreement with our results if the background density
is taken to be equal to a fixed fraction ($\approx0.44$) of the density at the spheroid surface.
Overall, this work makes progress toward developing simple time-dependent models of the CE phase,
for example by informing the development of drag force prescriptions for 1D spherically symmetric CE simulations,
which could be used to explore the parameter space of luminous red novae or in binary population synthesis studies.
\end{abstract}

\section{INTRODUCTION}
\label{sec:intro}
When a giant star engulfs a much smaller companion, 
the shared envelope drags on the companion and core, causing them to inspiral. 
This is known as a common envelope (CE) event and is important for a wide range of phenomena (see \citealt{Ivanova+20}, \citealt{Roepke+Demarco23}, and \citealt{Schneider+25} for recent reviews).
For stellar companions, the drag torque is dominated by gas dynamical friction, which has traditionally been modeled using Bondi-Hoyle-Lyttleton (BHL) theory \citep{Hoyle+Lyttleton39,Bondi+Hoyle44} and extensions thereof \citep{Dodd+Mccrea52,Ostriker99,Macleod+Ramirez-ruiz15a,Macleod+17,De+20}.

However, such models are unable to explain the drastic reduction in the gravitational drag torque 
seen to occur after the first periastron passage, 
during the slow spiral-in phase \citep[e.g.][]{Chamandy+19b,Chamandy+20}.%
\footnote{We note that `slow spiral-in' can have slightly different meanings in other works on CE evolution.}
This reduction happens partly because the binary separation becomes comparable to the BHL accretion radius,
causing the core and companion to experience a thrust from one another's high-density wakes 
that partially cancels the drag from their own wakes.
This effect has been discussed by \citet{Kim+08} (hereafter \citetalias{Kim+08}) 
in the idealized case of circular motion of equal point masses
whose 
effect on the background density $\rho_0$ is weak enough to be accurately described as a linear perturbation
in the variable $\lambda=(\rho-\rho_0)/\rho_0$.
\citetalias{Kim+08} built on earlier work 
that studied the drag on a single such ``perturber'' in a circular orbit \citep[e.g.][]{Kim+Kim07}.
Such studies make various assumptions and idealizations, not least that the linear treatment is adequate,
and thus it is not immediately clear to what extent they can be applied to the CE case.%
\footnote{We note that \citet{Sanchez-salcedo+Brandenburg01} 
studied the drag force in numerical simulations that solve a more general version of the equations  
and allow the orbit of the point mass to evolve.}
However, these models (or perhaps extensions thereof) could potentially be extremely useful,
so exploring the gap between such idealized models and global CE simulations may be a promising avenue. 
This is indeed one of the main goals of the present work (see~Section~\ref{sec:discussion_perturber}).

The scenario of a binary embedded in a gaseous medium also arises in other contexts, 
most famously that of binary supermassive black holes (BSMBHs), 
where gas dynamical friction may help to solve the ``last parsec problem''
(\citealt{Escala+04}; hereafter \citetalias{Escala+04}, \citealt{Escala+05,Dotti+07,Chapon+13,Tang+17,Li+Lai22,Williamson+22}), 
and compact object orbital tightening prior to coalescence, 
where it may help to solve the ``last astronomical unit problem'' \citep{Tagawa+18}.
In CE studies there seems to be an analogous ``last solar radius problem'' 
that prevents simulations from reaching final separations small enough 
to explain observations of post-CE binary systems \citep{Iaconi+Demarco19}.
This problem is likely caused, in part, by the torque evolution described above; 
long-duration, high-resolution CE simulations find that the orbit, does, in fact, 
continue to decay due to dynamical friction on long timescales \citep{Gagnier+Pejcha23,Chamandy+24}.
While it has been shown that extra physics like ionization and MHD can play an important role during the CE phase
\citep[e.g.][]{Ivanova+13a,Sand+20},
such effects are unlikely to have an order unity effect on the inspiral 
\citep[e.g.][]{Reichardt+20,Ondratschek+22,Chamandy+24}.
However, insufficient numerical resolution, including insufficiently small softening lengths, 
can cause the torque to be inaccurate, affecting the inspiral 
\citep{Iaconi+18,Chamandy+24,Gagnier+Pejcha25}.

Given the similarity between CE evolution and the other scenarios mentioned above, 
we explore how modelling approaches used in those cases might apply to the CE case.
For analysing their simulation of inspiralling equal-mass BSMBHs, 
\citetalias{Escala+04} developed a 
model for the dynamical friction torque that reproduces the evolution 
of the BSMBH separation at late times remarkably well, 
when the binary mass is much larger than the gas mass interior to the orbit. This condition  is generally also met in the slow spiral-in phase of CE evolution.  Their model takes advantage of the symmetry that arises when the components are of equal mass.
They model the torque as that applied by a constant-density, 
constant-aspect ratio prolate spheroid concentric with the binary but with major axis (in the orbital plane) 
lagging the axis passing through the binary components by a constant phase angle.
Inspired by \citetalias{Escala+04}, 
we try a similar approach for modelling the late-stage torque during CE
evolution and for this purpose we run a CE simulation with the companion mass equal to the core mass of the giant star.

This paper is organized as follows.
In Section~\ref{sec:methods} we describe the simulation methods.
In Section~\ref{sec:theory}, 
we summarize the idealized models to which our results are later compared,
namely the lagging uniform density prolate spheroid model of \citetalias{Escala+04}
and the circularly orbiting double-perturber model of \citetalias{Kim+08}.
In Section~\ref{sec:results}, we present results from our global CE simulation
and compare the torque measured from the simulation with predictions of the above models.
We discuss the successes and limitations of our work in Section~\ref{sec:discussion}.
Finally, we summarize and conclude in Section~\ref{sec:conclusions}.

\section{METHODS}\label{sec:methods} 
Our global hydrodynamic simulation is carried out with the 3D code \textsc{astrobear} 
\citep{Cunningham+09, Carroll-nellenback+13}. 
The setup is identical to the `AGB' run studied in \citet{Chamandy+20},
except that the mass of the binary companion particle is made to equal that of the AGB core particle.
The reader is referred to that paper for further details about the methods.
The AGB primary has initial mass $M_1 = 1.78 \Msun$ and radius $R_1 = 122.2 \Rsun$,
and the binary is initialized in a circular orbit with separation $a\init=124.0\Rsun$. 
The initial AGB density and pressure profiles are taken from a 1D single-star simulation
of a ZAMS $2\Msun$ star run using \textsc{mesa} \citep{Paxton+11,Paxton+13,Paxton+15,Paxton+19}.
Both the companion and core are modeled as gravitation-only point particles with mass $M_2=M\core=0.534\Msun$. 
The particles are not sink particles, i.e., they have fixed mass and do not accrete.
A spline function with a constant softening radius ($r\soft$) of $2.41 \Rsun$ 
is used to smooth the particle potential \citep{Hernquist+Katz89,Springel10}.
The 1D stellar profile inside $r=r\soft$ is modified to account for the presence of the AGB core particle
by solving a modified Lane-Emden equation and iterating over the particle mass until the correct interior mass
is recovered \citep{Ohlmann+17,Chamandy+18}.
Five levels of adaptive mesh refinement are employed and the regions around the particles
are resolved with a smallest cell size of $\delta_5=0.070\Rsun$ out to about $12\Rsun$ from each particle,
resulting in the ratio $r\soft/\delta_5=34.3$ cells per spline softening length.
The simulation domain is a cube of side $1150 \Rsun$, 
and the AGB star is placed at its centre, non-rotating in the lab frame.
The initial density ($\rho\amb$) and pressure ($P\amb$) of the ambient medium 
are set to $1.0 \times 10^{-9} \gcmcmcm$ and $1.1 \times 10^4 \dynecmcm$, respectively. 
The simulation is stopped at $t=299\da$, 
when the total energy (which should ideally be conserved) has increased by $6\%$
(see appendix~E of \citealt{Chamandy+24} for a discussion about energy conservation in our CE simulations).

The forces exerted on the primary core particle and the companion by gas are calculated using
\begin{equation}
\label{eq:f1}
    \bfF_1 = G M_{1,c} \int_V \rho(\bfs) \frac{\bfs - \bfs_1}{| \bfs - \bfs_1 |^3} \, \mathrm{d}^3 \bfs,
\end{equation}
\begin{equation}
    \label{eq:f2}
    \bfF_2 = G M_2 \int_V \rho(\bfs) \frac{\bfs - \bfs_2}{| \bfs - \bfs_2 |^3} \, \mathrm{d}^3 \bfs\\
\end{equation}
 where $\rho(\mathbf{s})$ is density of gas at position $\mathbf{s}$, 
$V$ is the volume of the whole simulation box and $\mathbf{s}_1$ and $\mathbf{s}_2$ are the position vectors of the core and the companion, respectively. 
The $z-$component of the torque about the particle centre of mass is calculated using
\begin{equation}
\label{eq:sim_torque}
    \tau_z = \frac{a}{m}\left(M_2 \bfF_{1,\phi} + M\core \bfF_{2,\phi}\right),
\end{equation}
where $a$ is the particle separation projected onto the $x$-$y$ ($r$-$\phi$) plane, 
which remains almost exactly parallel to the orbital plane of the particles during the simulation,
and $m=M\core+M_2$ is the combined mass of the binary point particles. 
In this work, $M\core=M_2=m/2$.
The separation $a$ and orbital evolution are shown in Figure~\ref{fig:separation}.
The quantity $-\tau_z$ (where the minus sign is introduced to cancel the minus sign resulting from net drag
so that we can plot a positive quantity) 
is shown as a solid black line in both panels of Figure~\ref{fig:torque}.

\section{IDEALIZED MODELS FOR COMPARISON WITH SIMULATION RESULTS}
\label{sec:theory} 

Our first method of approximation is to model the gravitational torque on the particles as that exerted 
by a uniform ellipsoid rotating at the instantaneous orbital angular frequency, with its major axis
situated in the orbital plane and lagging the line joining the particles by a phase angle.

\subsection{Uniform density lagging prolate spheroid model} \label{sec:spheroid}
The gravitational potential at a point $(x,y,z)$ inside an ellipsoid of uniform density $\rhobar$ is given by
\begin{equation}\label{eq:potential}
     \Phi = 
     \pi G \rhobar \left(\alpha x^2 + \beta y^2 + \zeta z^2 - \chi \right),
\end{equation}
where the coefficients $\alpha$, $\beta$, $\zeta$ and $\chi$ are given in \citet[p.~146-154, 700]{lamb1879}, 
\citet[p.~38-45]{Chandrasekhar69}, and \citet[p.~83-95]{Binney+Tremaine08}. 
Take the origin to be the particles' orbital centre of mass (CM) in the simulation and the $x$-axis to be the axis passing through both particles. 
For the special case of a constant-density prolate spheroid with major axis lagging this binary axis by the angle $\Delta\phi$, 
the $z$-component of torque at any point inside the spheroid is given by
\begin{equation}\label{eq:torque_expression1}
    \begin{aligned}
        \tau_z = \pi G \rhobar m 
        &\Big\{\sin \left(2\Delta \phi\right) 
        \left(\beta - \alpha\right)\left(y^2-x^2\right) \\
        &+ 2xy\left[\alpha\cos\left(2\Delta\phi\right) 
        - \beta\cos\left(2\Delta\phi\right)\right] \Big\},
    \end{aligned}
\end{equation}
where
\begin{equation}
    m = M\core + M_2 = 1.068\Msun,
\end{equation}
\begin{equation}
    \alpha = \frac{1-e^2}{e^3} \ln \frac{1+e}{1-e}-2 \frac{1-e^2}{e^2},
\end{equation}
\begin{equation}
        \beta = \frac{1}{e^2}-\frac{1-e^2}{2 e^3} \ln \frac{1+e}{1-e},
\end{equation}
\begin{equation}\label{eq:eccentricity}
        e =\sqrt{1-\Btilde^2 / \Atilde^2},
\end{equation}
$\Atilde$ and $\Btilde$ are the semi-major and semi-minor axes of the spheroid, respectively,
and $e$ is the spheroid eccentricity.
As a consequence of Newton's third theorem \citep[][p.~87]{Binney+Tremaine08}, 
the ellipsoid’s mass outside the similar, 
concentric ellipsoid (of the same orientation) 
that passes through the particles exerts no net force on the binary. 
Therefore, the torque does not depend directly on the size of the ellipsoid.
However, below we obtain $\rhobar$ by averaging the density inside an ellipsoidal contour, 
so the overall size of the contour does play a role.
In our choice of coordinate system, 
the expression for the aggregate torque on both particles reduces to 
(\ref{sec:torque_derivation})
\begin{equation}\label{eq:torque_z}
    \tau_z = - \frac{1}{2} \pi G m \left( \beta - \alpha \right) \cos (\Delta \phi) \sin (\Delta \phi) \rhobar a^2,
\end{equation}
where $a$ is the orbital separation.

\subsection{Simplified double-perturber model} \label{sec:kim}
Another approach is to solve the governing equations after making various approximations.
\citetalias{Kim+08} solved numerically a linearized version of the hydrodynamics equations for two perturbers in a circular orbit, 
building on a study of a single perturber in a circular orbit, \citet{Kim+Kim07}, 
which in turn extended the work of \citet{Ostriker99} for rectilinear motion.
The \citetalias{Kim+08} model is valid if the perturbers are weak and if various complicating effects -- 
gas self-gravity, orbital eccentricity,
the density gradient and rotation of the gas in which the perturbers are embedded, 
the motion of the binary CM,
and the time dependence due to orbital evolution -- can all be safely neglected.
Their model further assumes an ideal gas with adiabatic index $\gamma=5/3$,
which is also assumed in our CE simulation.

Below we restrict our discussion to the case of equal mass perturbers. 
While \citetalias{Kim+08} also focuses on this case, 
we note that their general model includes the mass ratio as a parameter.
The force on each perturber can be written as 
\begin{equation}\label{F_K08}
  \vec{F} = -\mathcal{F}(\mathcal{I}_R\bfRhat + \mathcal{I}_\phi\bfphihat), 
\end{equation}
where $\mathcal{I}_R\bfRhat$ and $\mathcal{I}_\phi\bfphihat$ characterize the radial and azimuthal components in the orbital plane,
\begin{equation}\label{mathcalF}
  \mathcal{F} \equiv \frac{4\pi\rho_0 (GM\pert)^2}{V\pert^2},
\end{equation}
$M\pert$ and $V\pert$ are the mass and speed of a perturber, 
and $\rho_0$ is the unperturbed gas density. 
Further, the drag force is exerted both by the wake of the perturber and by the wake of its companion,
so we can write
\begin{equation}\label{I_R}
  \mathcal{I}_R = \mathcal{I}'_{R} + \mathcal{I}''_{R} 
\end{equation}
and
\begin{equation}\label{I_phi}
  \mathcal{I}_\phi = \mathcal{I}'_{\phi} + \mathcal{I}''_{\phi},
\end{equation}
where the first term, denoted by prime, is due to the perturber's own wake and the second term, 
denoted by double-prime, is due to the wake of the other perturber.

\citetalias{Kim+08} (see also \citealt{Kim+Kim07}) solved numerically for the drag force exerted on the perturbers 
by their wakes and obtained fitting formulas that closely match the numerical solutions.
The adiabatic sound speed is given by 
\begin{equation}
  c\sound = \left(\gamma\frac{P}{\rho}\right)^{1/2},
\end{equation}
where $\gamma=5/3$ and $P$ is the gas pressure.
The sonic Mach number is given by
\begin{equation}\label{Mach}
  \Mach\pert = \frac{V\pert}{c_\mathrm{s,0}}, 
\end{equation}
where $c_\mathrm{s,0}$ is the sound speed in the unperturbed medium.
We shall see below that the subsonic regime, with $\Mach\pert<1$, is most relevant for this work.
In this regime the \citetalias{Kim+08} fitting formulas are
\begin{equation}
  \mathcal{I}'_{\phi}\label{Iphi_prime}
  = 0.7706\ln\left(\frac{1+\Mach\pert}{1.0004 - 0.9185\Mach\pert}\right) - 1.4703\Mach\pert
\end{equation}
and
\begin{equation}\label{Iphi_doubleprime}
  \mathcal{I}''_{\phi}
  = -0.022\Mach\pert^2(10-\Mach\pert)\tanh\left(\frac{3\Mach\pert}{2}\right).
\end{equation}
An analytical solution has been derived by \citet{Desjacques+22} that agrees with the numerical solutions of \citetalias{Kim+08},
but as the analytical solution is cumbersome 
we choose to compare our solutions with the fitting formulas~\eqref{Iphi_prime} and \eqref{Iphi_doubleprime}.%
\footnote{See also \citet{Stahler+10} for an alternative analytic solution to the problem.}

The $z$-component of the torque on a single perturber in the rest frame of the CM of the perturbers is given by
$-(a/2)\mathcal{F}\mathcal{I}_\phi$,
where $\mathcal{F}$ is given by equation~\eqref{mathcalF} and $\mathcal{I}_\phi$ by equation~\eqref{I_phi}.
Thus, for equal mass perturbers with equal Mach numbers, the torque on the binary is given by
\begin{equation}\label{torque_Kim08}
  \tau_z = -a\mathcal{F}\mathcal{I}_\phi.
\end{equation}

However, in the simulation the azimuthal speeds of the equal-mass point particles are not precisely equal
due to asymmetry in the gas distribution.
One can generalize equations~\eqref{Iphi_prime} and \eqref{Iphi_doubleprime} 
by replacing $V\pert$ with $V_{i,\phi}$ in equation~\eqref{mathcalF},
and $\Mach\pert$ with 
\begin{equation}\label{Mach_i}
  \Mach_i = V_{i,\phi}/c_\mathrm{s,0}
\end{equation}
in equations~\eqref{Iphi_prime} and \eqref{Iphi_doubleprime}, 
where $i$ represents the particle on which the force is being calculated,
i.e.~$1$ for the AGB core particle and $2$ for the companion particle.
In this case, the total torque on the binary is given by
\begin{equation}\label{torque_Kim08_general}
  \tau_z = -\frac{a}{2}\left(\mathcal{F}_1\mathcal{I}_{1,\phi} + \mathcal{F}_2\mathcal{I}_{2,\phi}\right).
\end{equation}
Writing this general torque equation for equal mass perturbers explicitly, we have
\begin{equation}\label{torque_Kim08_general_explicit}
\begin{split}
  \tau_z &= -2\pi a\rho_0(GM\pert)^2 \\
  &\quad\times\left[
  \frac{I'_\phi(\Mach_1)+I''_\phi(\Mach_1)}{V_{1,\phi}^2}
  +\frac{I'_\phi(\Mach_2)+I''_\phi(\Mach_2)}{V_{2,\phi}^2}
  \right].
\end{split}
\end{equation}

Equations~\eqref{eq:torque_z} and \eqref{torque_Kim08_general_explicit} 
represent the approximate models for the torque on the binary 
that we will use to compare with the simulation.

\section{RESULTS}
\label{sec:results}

\begin{figure}
    \centering
    \includegraphics[width=\columnwidth, clip = true,trim={7 0 0 0}]{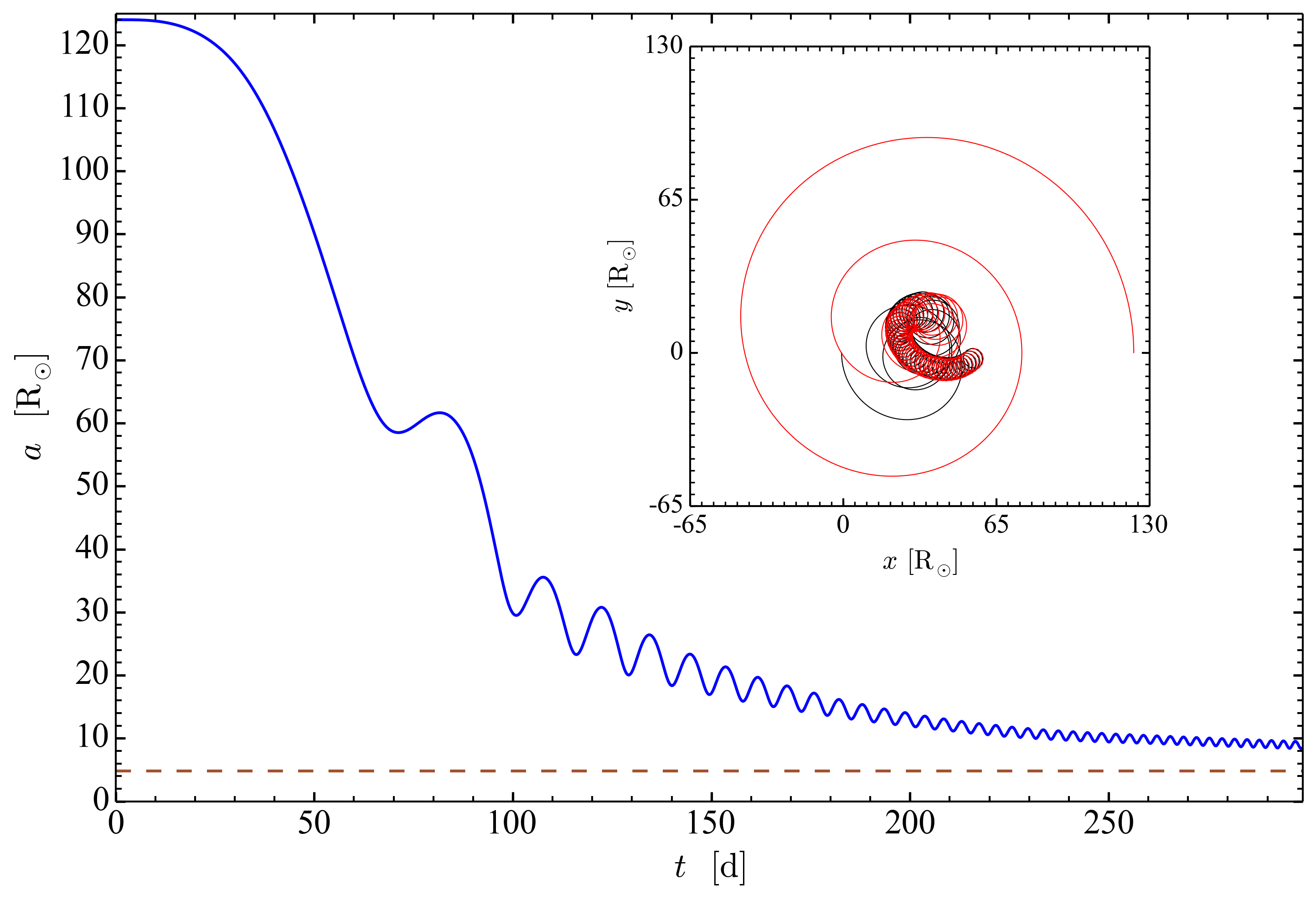}
    \caption{Separation between the AGB core and companion particles as a function of time.
    The dashed line shows twice the softening radius, $2r\soft$,
    and the inset shows the orbit of the primary core (black) and companion (red).}
    \label{fig:separation}
\end{figure}

\begin{figure*}
    \centering
    \includegraphics[width=1\textwidth]{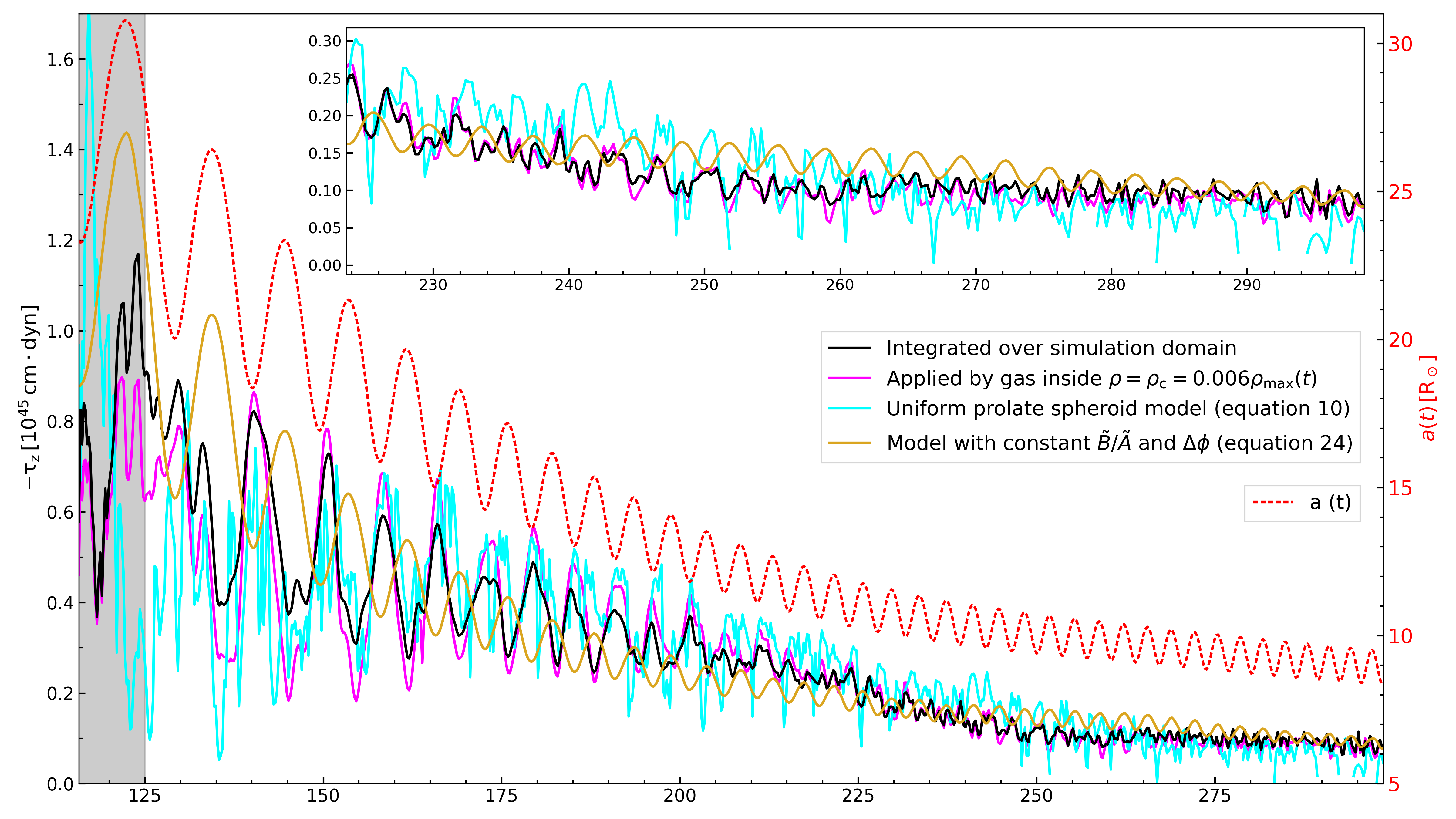}
    \includegraphics[width=1\textwidth]{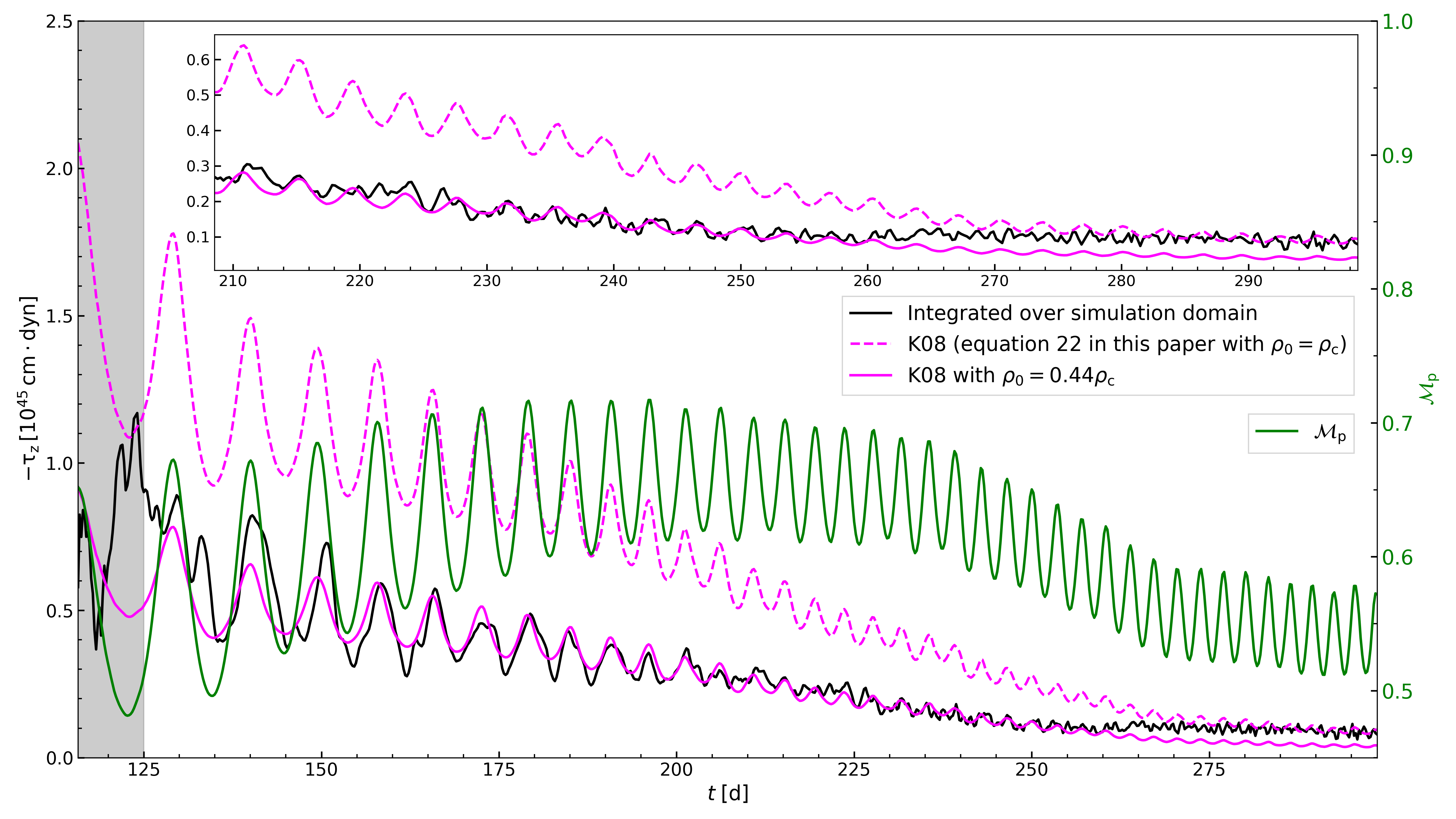}
    \caption{
    The $z$-component of torque on the binary system about the particle CM (left axis). 
    \textit{Top}: Shown is the torque (i)~measured directly from the simulation (black), 
    (ii)~including only contributions out to the contour $\rho=\rho\crit=0.006\rho\ma(t)$ (magenta), 
    (iii)~reconstituted using equation \eqref{eq:torque_z} with $\rhobar$, 
    $\Delta\phi$ and $\Btilde/\Atilde$ measured from the simulation (cyan), 
    (iv)~reconstituted using equation \eqref{eq:torque_z_mean} for a co-rotating spheroid 
    with $\langle\Delta\phi\rangle = 14.9^\circ$ and $\langle\Btilde/\Atilde\rangle=0.654$ (orange).
    The orbital separation of the particles, $a$, is shown on the right axis (dashed red).
    The inset shows a zoom-in of the torque at late times. 
    \textit{Bottom}: The $z$-component of the torque (i)~measured directly from the simulation 
    (same as in the top panel, black), 
    (ii)~calculated from equation \eqref{torque_Kim08_general_explicit} 
    with $c_\mathrm{s,0}$ taken as the mean sound speed $\overline{c}\sound$
    inside the surface $\rho=\rho\crit$
    and $\rho_0$ taken as the density on this surface (dashed magenta), 
    and (iii)~the same but now $\rho_0$ is taken to be $0.44$ times the value on the surface (solid magenta).
    The particle Mach number $\Mach_\mathrm{p}$,
    obtained by dividing the particle tangential speed in the particle CM frame 
    by $\overline{c}\sound$, is shown on the right axis in solid green.
    }
    \label{fig:torque}
\end{figure*}

\begin{figure*}
    \centering
    \includegraphics[width=0.46\textwidth]{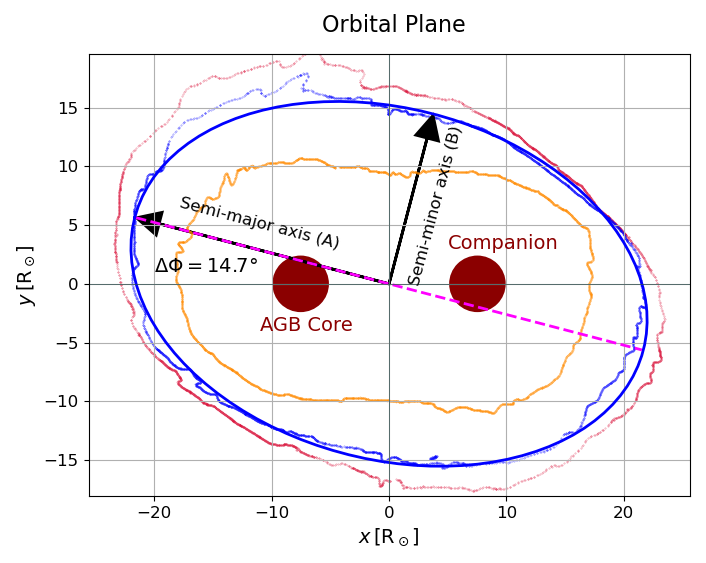}
    \includegraphics[width=0.46\textwidth]{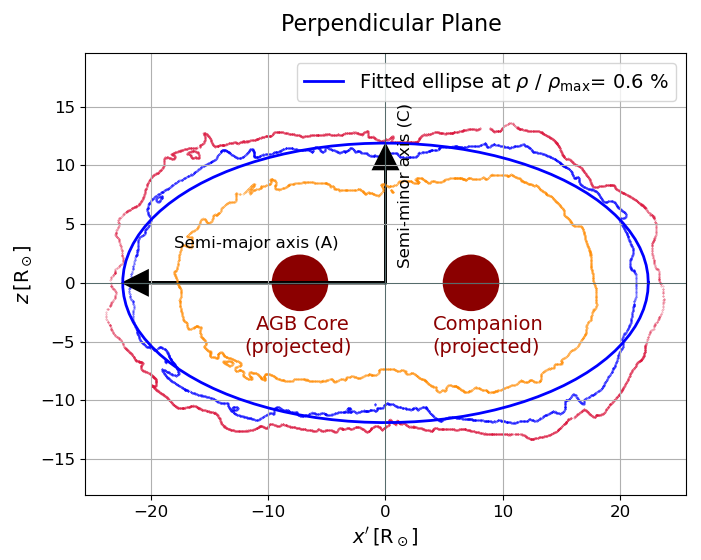}
    \includegraphics[width=0.45\textwidth]{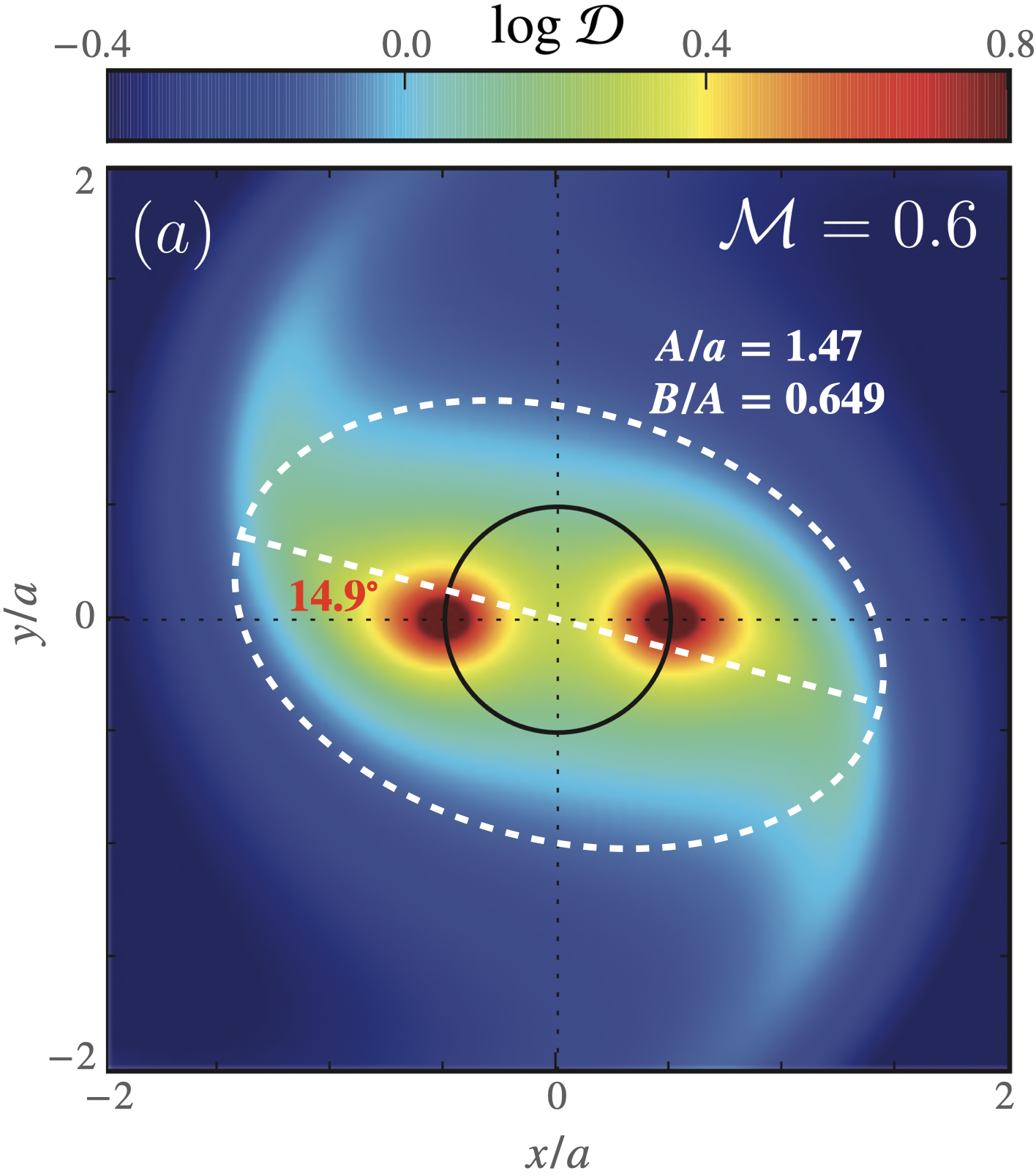}
    \includegraphics[scale=0.40]{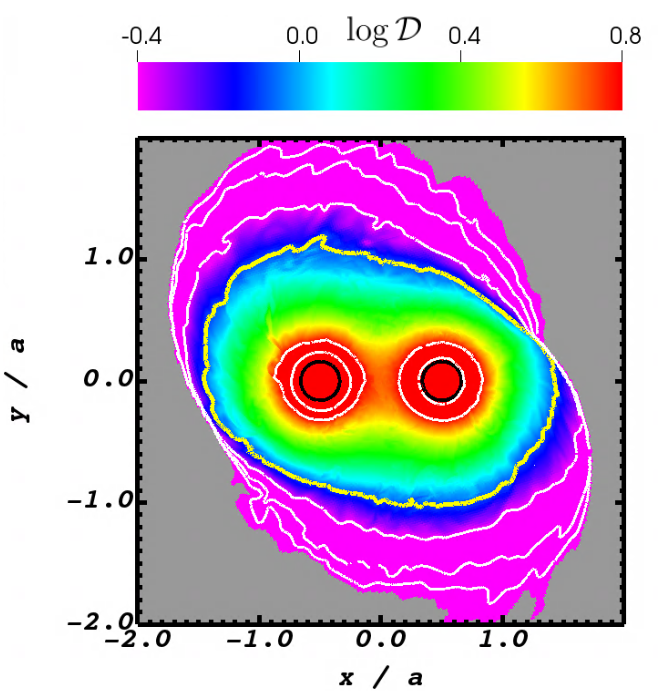}
    \caption{
    \textit{Top left}:~Density contours at $\rho=0.01\rho\ma(t)$, $0.006\rho\ma(t)$, 
    and $0.005\rho\ma(t)$ in the orbital plane at the time $t=188.7\da$, 
    with ellipse fitted to the contour $\rho=\rho\crit=0.006\rho\ma(t)$, which was found to enclose effectively all of the gas producing significant torque (see also Figure~\ref{fig:torque}). The ellipse is phase-shifted by an angle $\Delta\phi$ with respect to the axis that passes through the particles. 
    \textit{Top right}:~Similar to the top-left panel but now showing the plane perpendicular to the orbital plane 
    and rotated clockwise by the angle $\Delta\phi$ about the orbital axis,
    shown by the dashed line in the top left panel.
    The length of the ellipse major axis is set equal to that in the orbital plane, 
    but the length of the minor axis is allowed to differ.
    \textit{Bottom left}:~Adapted from figure~1 of \citetalias{Kim+08}, 
    showing the steady state for $\Mach\pert=0.6$ sliced through the orbital plane in their idealized double-perturber model.
    The black circle shows the orbit of the point masses that perturb the background density.
    Colour shows the density contrast $\log\mathcal{D}$, 
    where $\mathcal{D}=(c_\mathrm{s,0}^2a/Gm)\lambda$ with $m$ the binary mass and $\lambda=(\rho-\rho_0)/\rho_0$.
    Overplotted for comparison is the time-averaged best fit ellipse in the orbital plane from our simulation, 
    with parameter values noted on the plot (see also Figure~\ref{fig:params}).
    \textit{Bottom right}:~Similar to the bottom left panel but now showing $\log\mathcal{D}$ for our simulation, 
    at the same time as the top row, when $\rho_0=0.44\rho\crit=3.16\times10^{-6}\gcmcmcm$,
    and $c_\mathrm{s,0}=\overline{c}\sound=93.0\kms$.
    The region outside $0.44\rho\crit$ has negative values of $\mathcal{D}$ and is coloured grey.
    The contour $\rho=\rho\crit$ is plotted in yellow.
    The white contours near the softening radius (black circles, 
    AGB core on the left and companion on the right) show contours of $\log\mathcal{D}=1.2\,\mathrm{and}\,0.8$, 
    while the contours outside $\rho=\rho\crit$ show $\log\mathcal{D}=-0.4, -0.8\,\mathrm{and}\,-1.2$.
    }
    \label{fig:fitting}
\end{figure*}

\begin{figure}
    \centering
    \includegraphics[width=\linewidth]{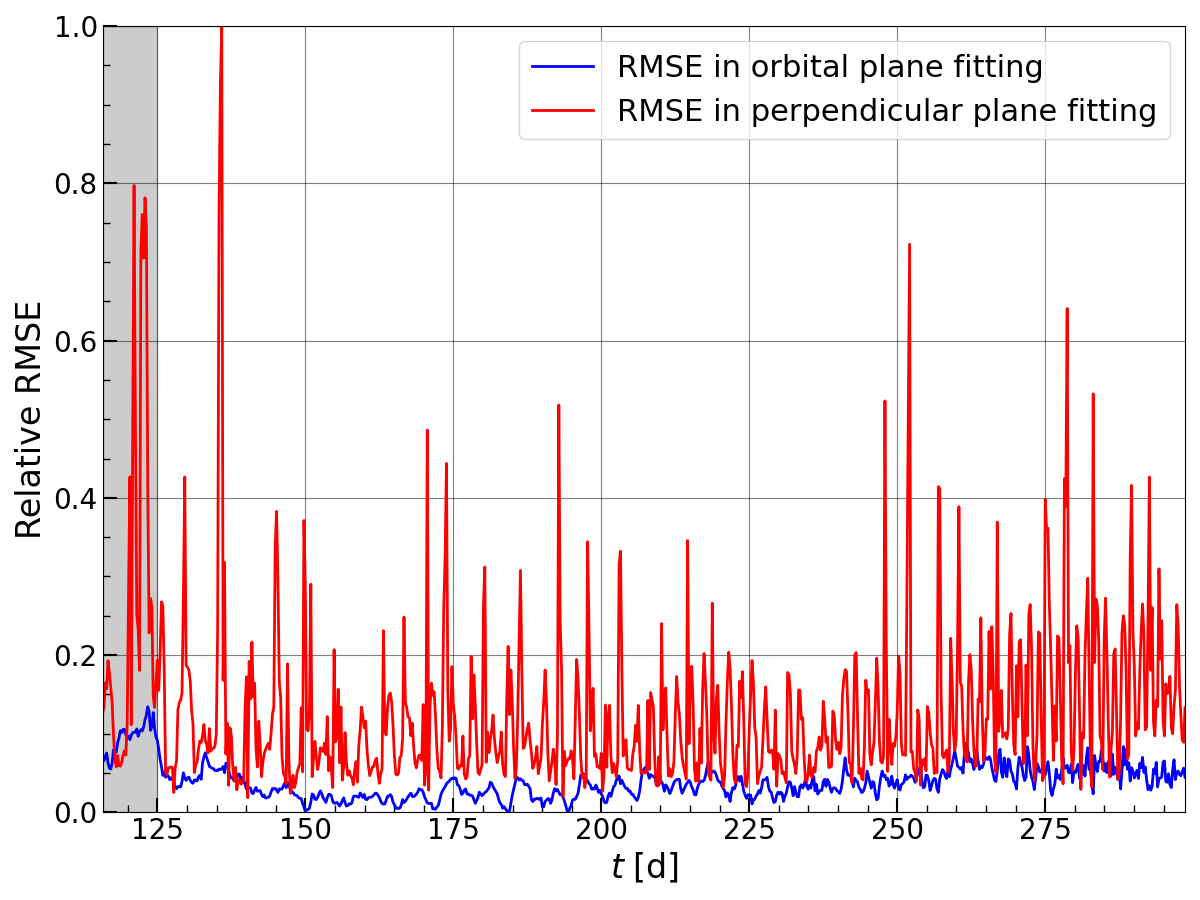}
    \caption{
    Relative root mean squared error (RMSE) of ellipse fitting in both planes for the ellipsoid model discussed in Section~\ref{sec:spheroid}.
    (the highest value is set to unity and other values are calculated with respect to it).
    We observe a higher RMSE in the perpendicular plane as the major axis in this plane 
    is forced to have the same value as that in the orbital plane.
    We start our analysis at $t=125\da$.
    }
    \label{fig:rmse_evol}
\end{figure}

\begin{figure*}
    \centering
    \includegraphics[width=1\textwidth]{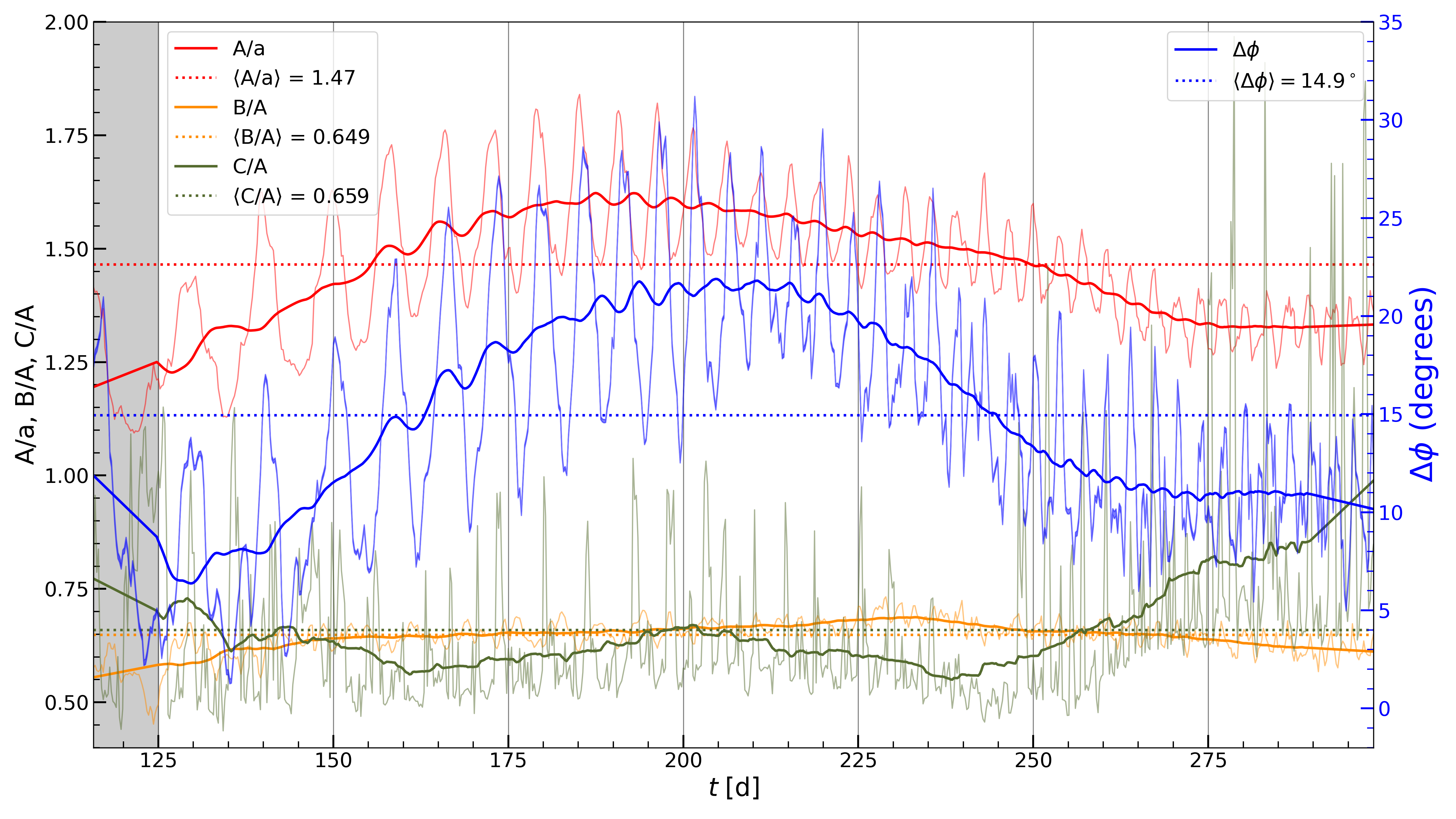}
    \caption{Time evolution of key fit parameters for the lagging spheroid model:
    (i)~ratio of semi-major axis $A$ 
    to separation $a$,
    (ii)~ratios of semi-minor axis ($B$ in the orbital plane and $C$ in the perpendicular plane) 
    to semi-major axis $A$, and
    (iii)~lag angle $\Delta \phi$ between the binary axis and the major axis of the fitted ellipsoid (right axis). 
    The simulation output (thin lines) oscillates rapidly with time.
    Thick lines show $10\da$-moving averages and the dotted lines 
    show mean values over the time domain 
    of the analysis (125 days onward).
    }
    \label{fig:params}
\end{figure*}

\begin{figure}
    \centering
    \includegraphics[width=\linewidth]{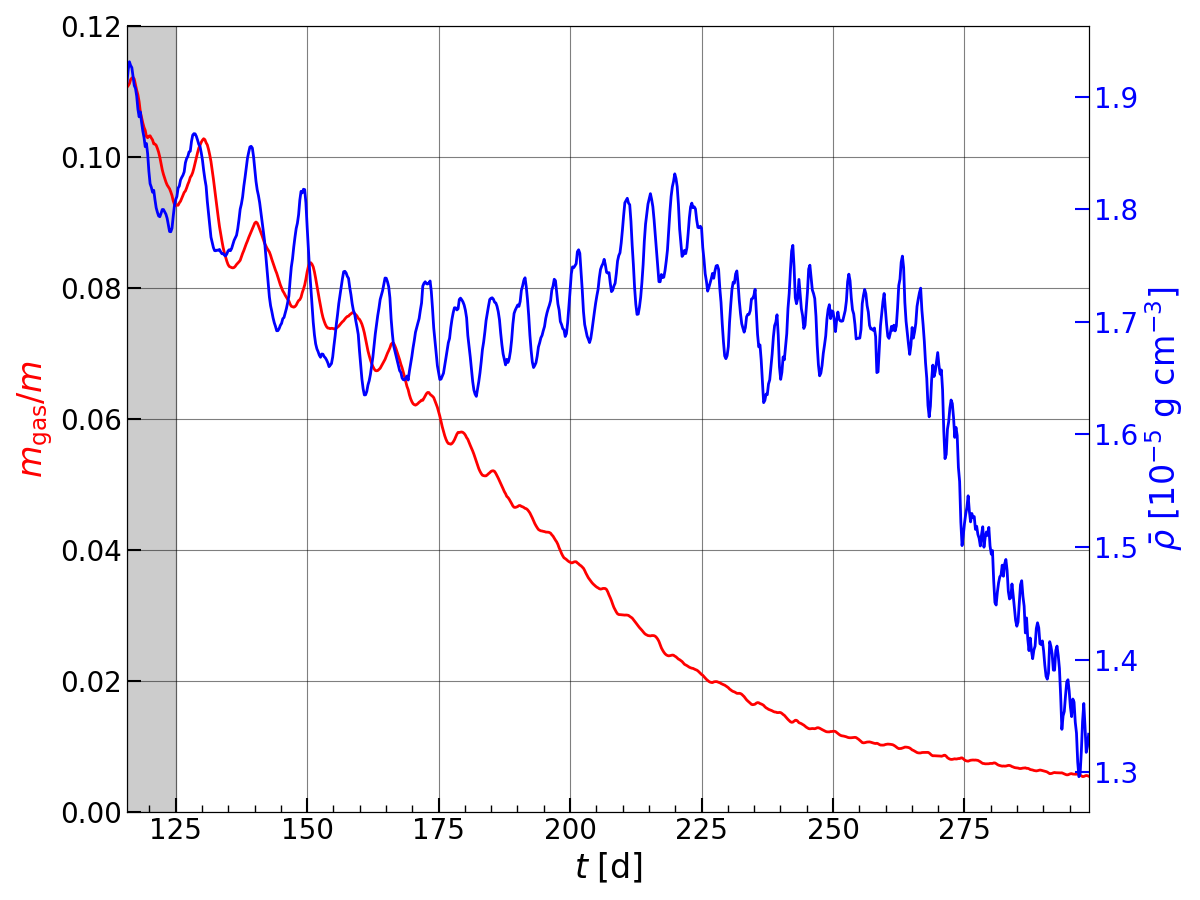}
    \caption{Time evolution of the ratio of the gas mass enclosed by the equipotential surface $\rho=\rho\crit=0.006\rho\ma(t)$ 
    to the binary particle mass $m= M\core + M_2$ (left axis) and the mean density inside that surface (right axis).
    }
    \label{fig:mass}
\end{figure}

\begin{figure*}
    \centering
    \includegraphics[scale=0.17, trim=0 0 0 0, clip=True]{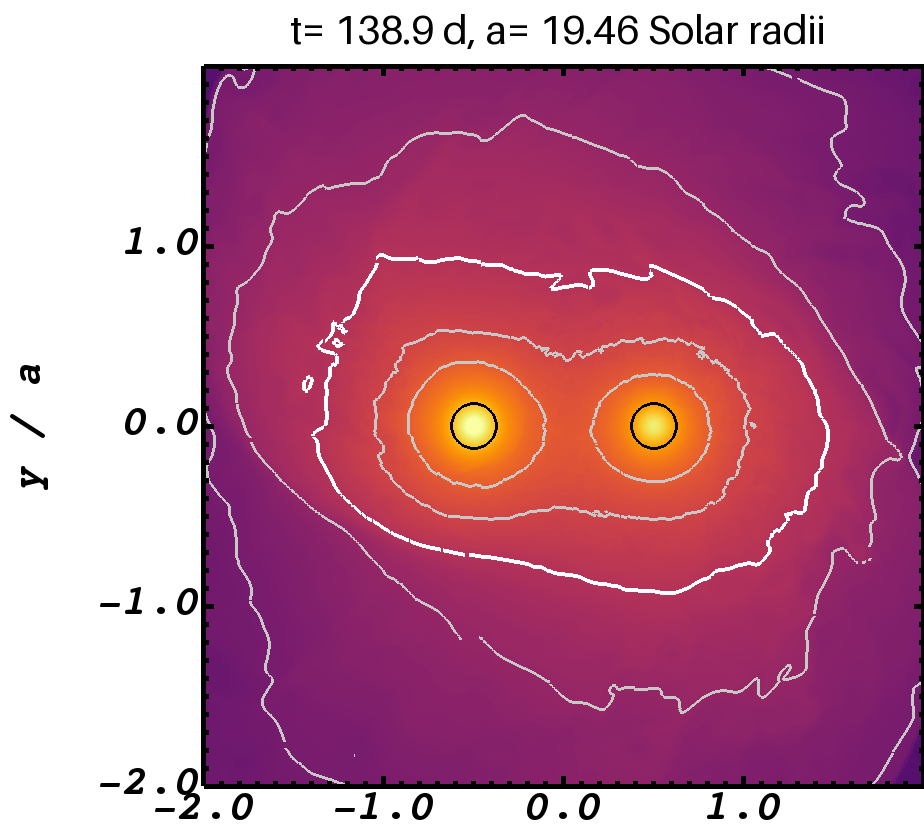}
    \includegraphics[scale=0.17, trim=0 0 0 0, clip=True]{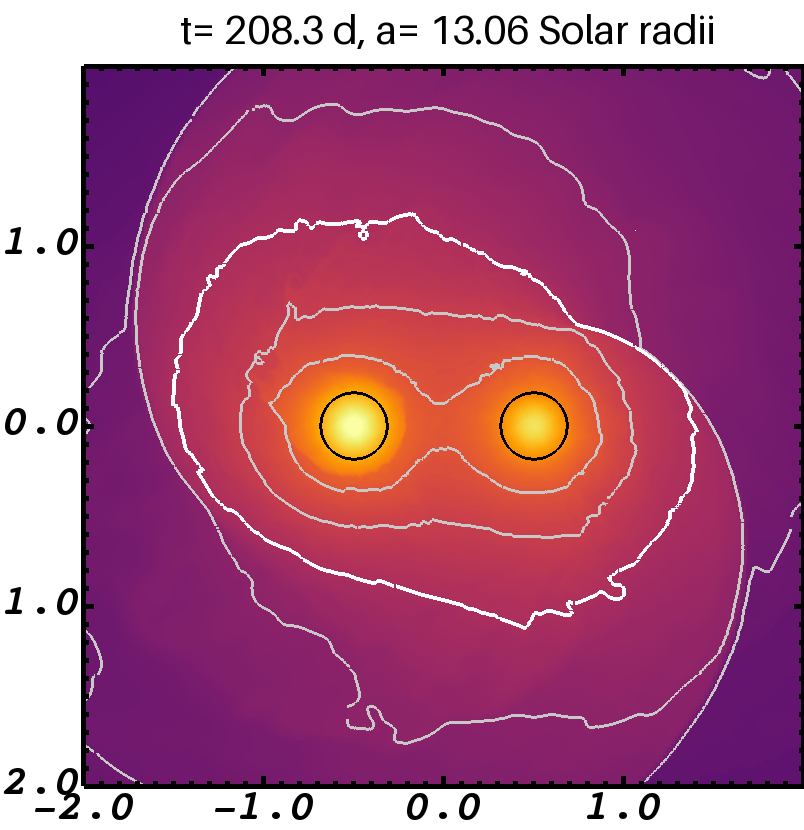}
    \includegraphics[scale=0.199, trim=0 0 0 0, clip=True]{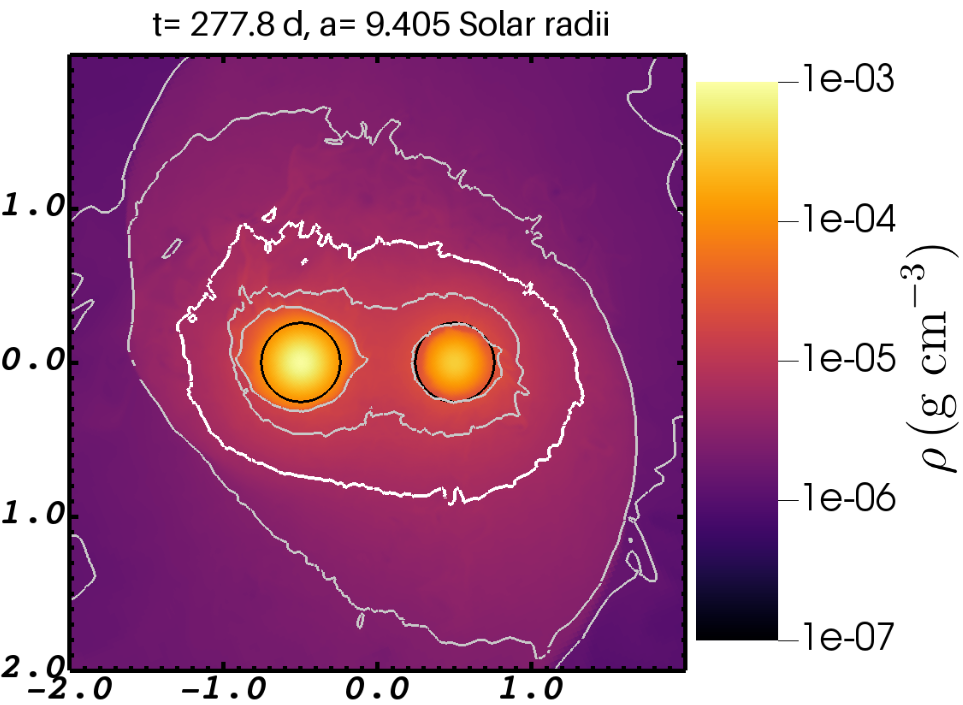}\\
    \includegraphics[scale=0.17, trim=0 0 0 0, clip=True]{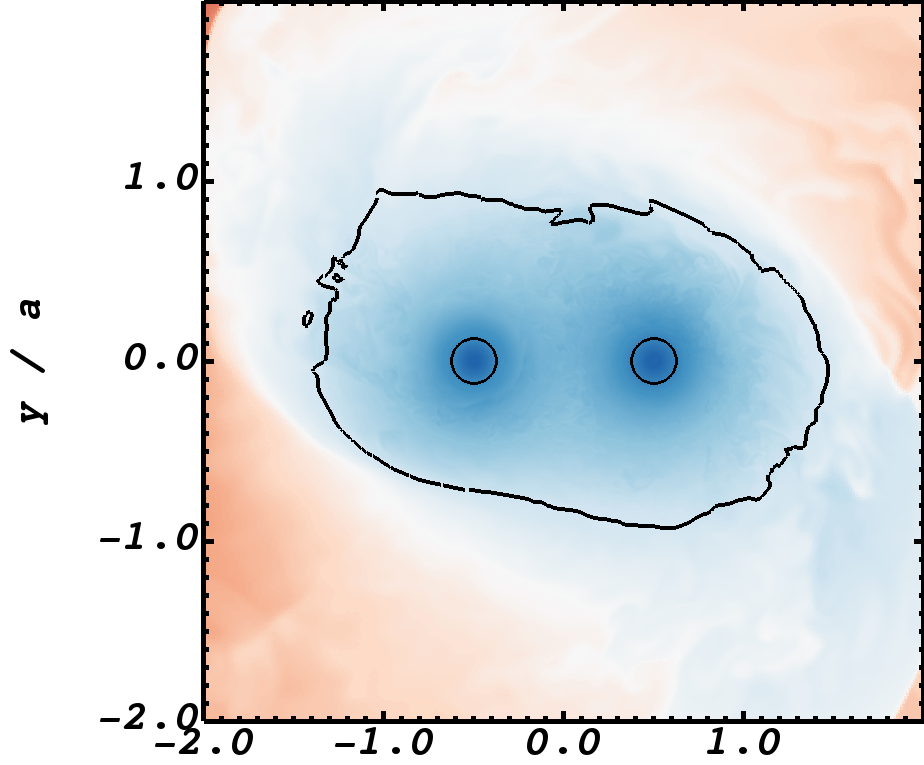}
    \includegraphics[scale=0.17, trim=0 0 0 0, clip=True]{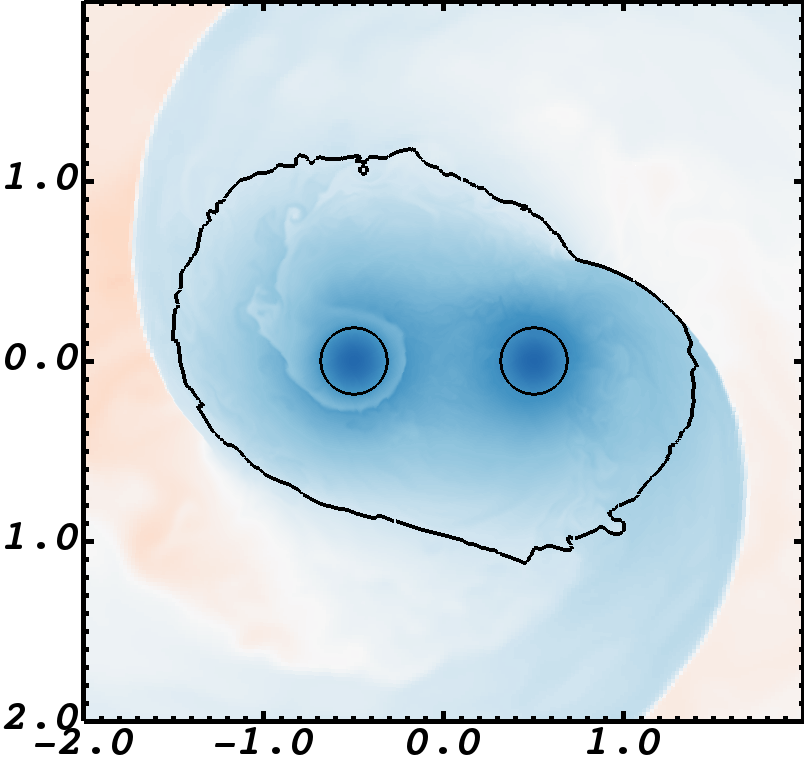}
    \includegraphics[scale=0.228, trim=0 0 0 0, clip=True]{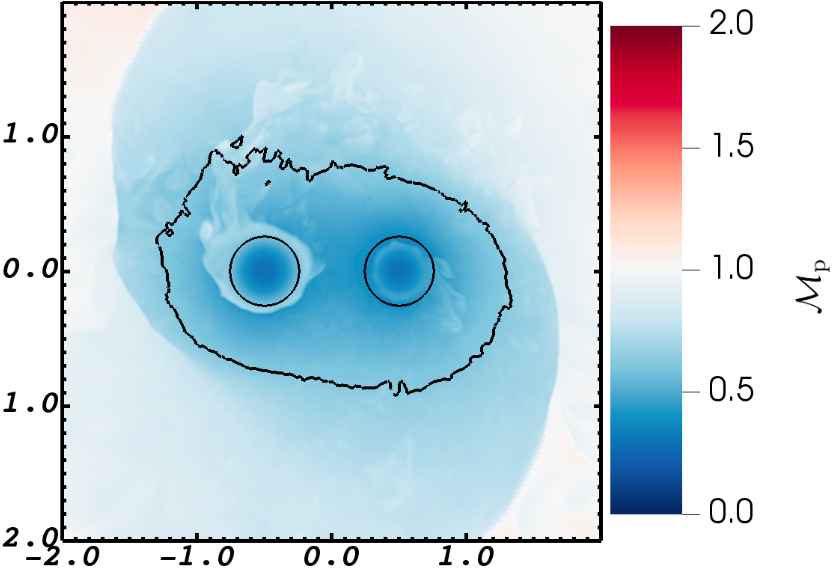}\\
    \includegraphics[scale=0.17, trim=0 0 0 0, clip=True]{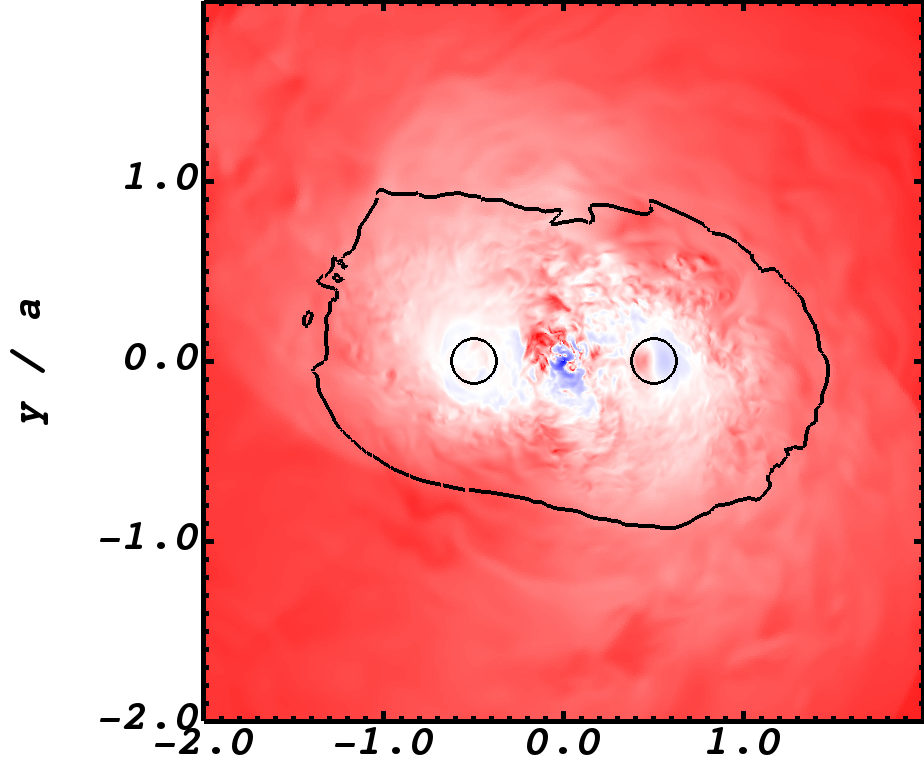}
    \includegraphics[scale=0.17, trim=0 0 0 0, clip=True]{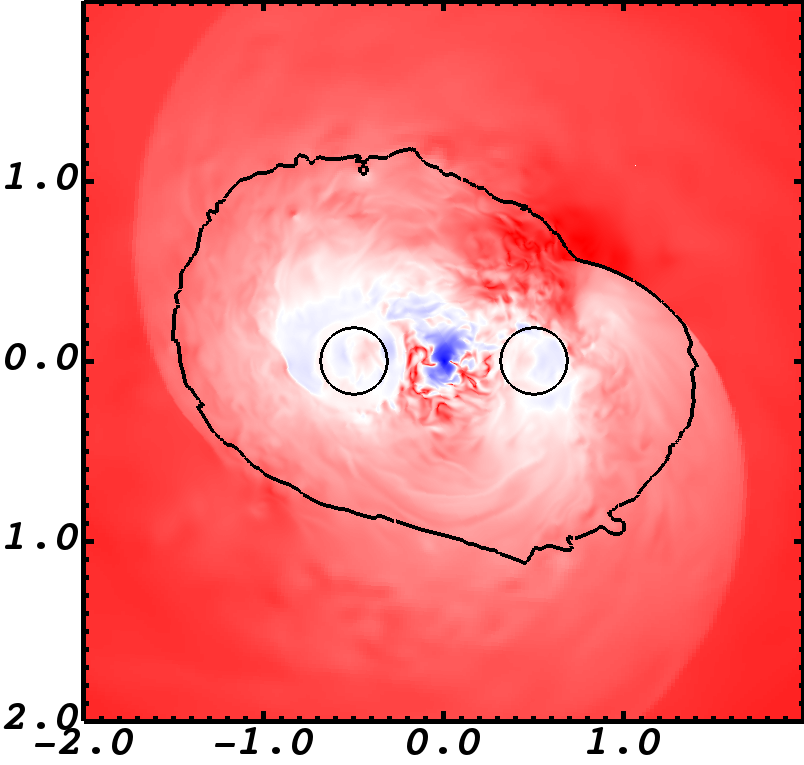}
    \includegraphics[scale=0.198, trim=0 0 0 0, clip=True]{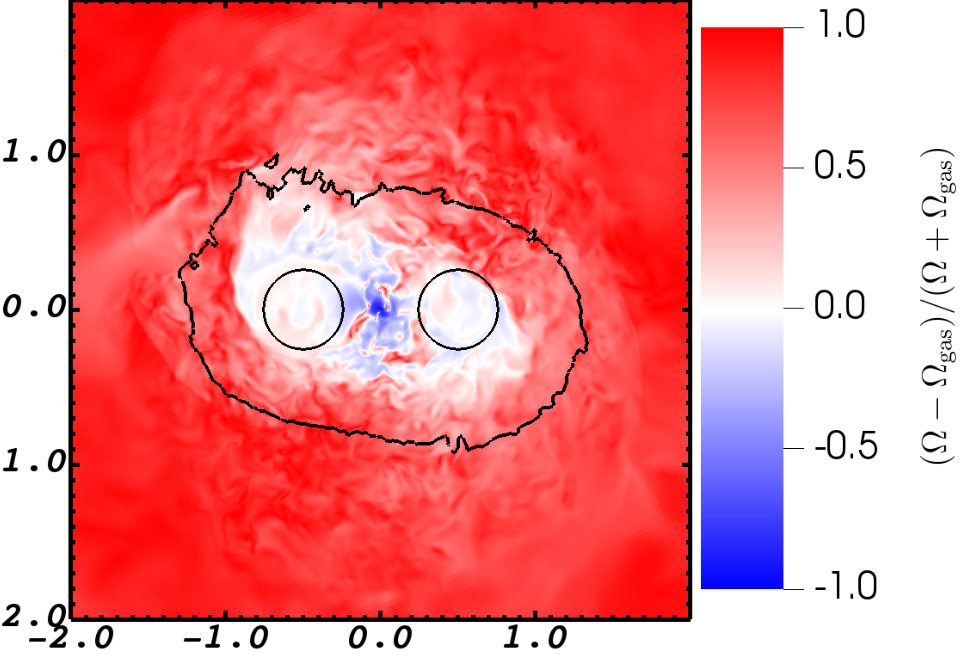}\\
    \includegraphics[scale=0.17, trim=0 0 0 0, clip=True]{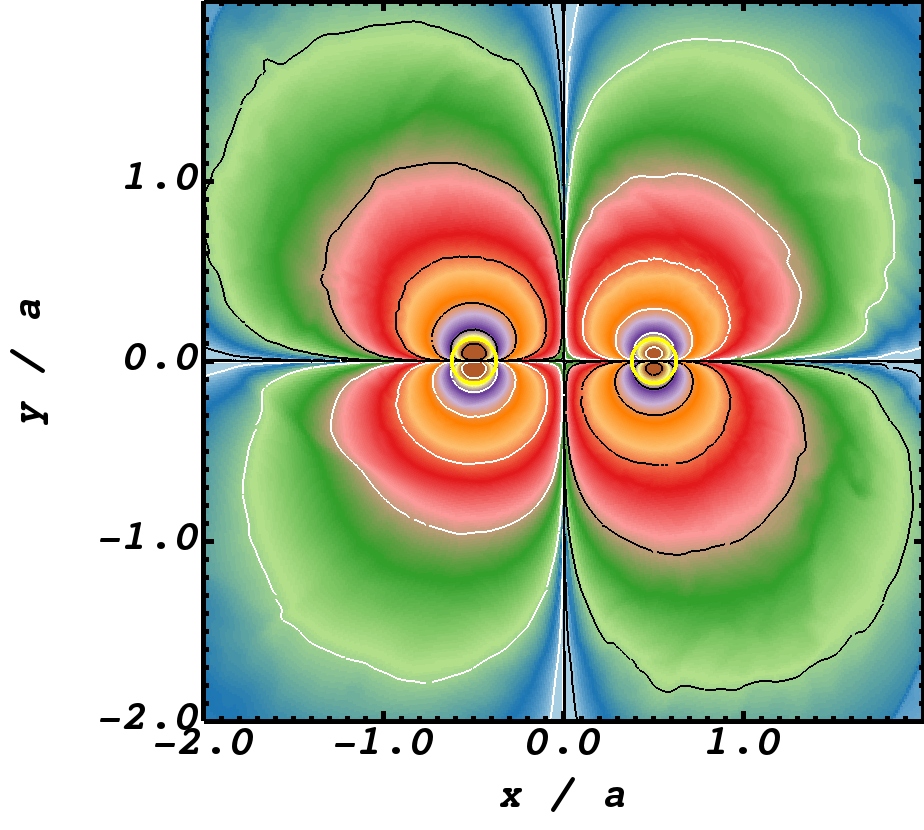}
    \includegraphics[scale=0.17, trim=0 0 0 0, clip=True]{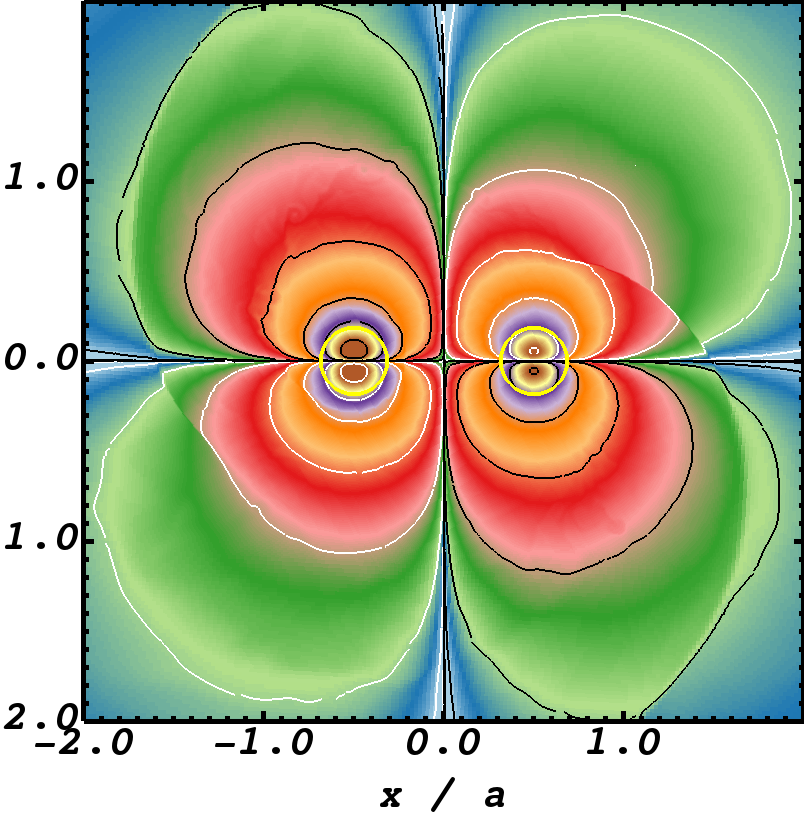}
    \includegraphics[scale=0.228, trim=0 0 0 0, clip=True]{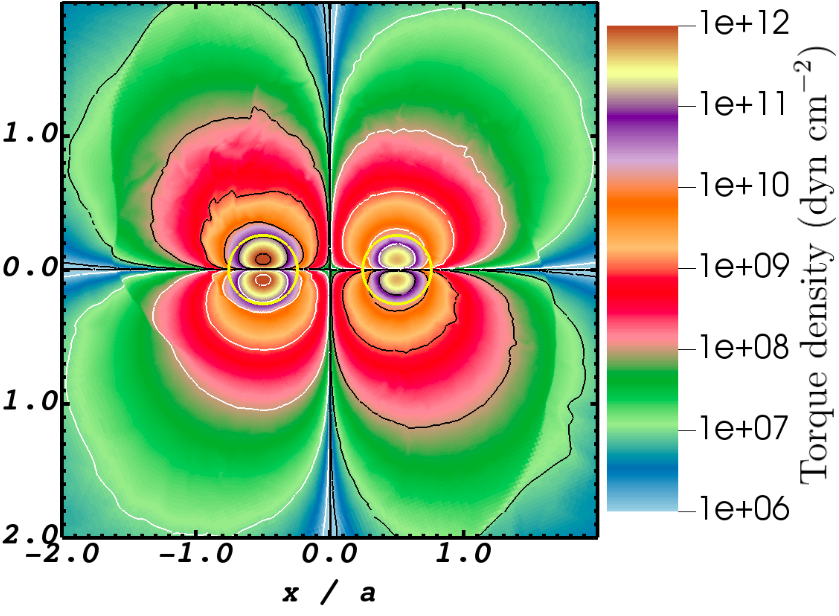}
    \caption{From left to right, columns show snapshots in the orbital plane at times $t=138.9$, $208.3$, and $277.8$ days. 
    Primary core (left) and companion (right) softening spheres are shown in black or yellow in the bottom row. 
    Rows from top to bottom are: (i)~Mass density ($\rho$, thick white contour for $\rho=\rho\crit$, 
    and thin grey contours showing $\rho=4\rho\crit, 2\rho\crit, 0.5\rho\crit\,\text{and}\, 0.25\rho\crit$);
    (ii)~The local Mach number of the 
    particles
    $V_\phi/c\sound$, 
    where $V_\phi$ is the $\phi$-component 
    of the particle velocity in the particle centre of mass frame
    ($V_{1,\phi} = V_{2,\phi}$) 
    and $c\sound$ is the sound speed, which depends on position (blue 
    denotes the subsonic region and red 
    the supersonic region); 
    (iii)~The difference over the sum of the angular speeds of the perturbers and gas $(\Omega-\Omega\gas)/(\Omega+\Omega\gas)$ 
    (red being the region dominated by $\Omega$, blue being the region dominated by $\Omega\gas$);
    (iv)~The torque density on both particles. Black and white contours 
    respectively show the negatively and positively contributing regions to the $\phi$-component of the drag.
    The highest contour levels near the particles are $\pm10^{12}\dyncmcm$, 
    and contours are also plotted for $\pm10^{11}\dyncmcm$, $\pm10^{10}\dyncmcm$, and so on.
    Movies starting from $t=115.7$ days to the end of the simulation are available at \href{https://doi.org/10.5281/zenodo.17575148}{https://doi.org/10.5281/zenodo.17575148}.
    }
    \label{fig:snapshots}
\end{figure*}

\subsection{Orbit}
\label{sec:orbit}
The evolution of the distance $a$ between the AGB core particle and companion is shown in Figure~\ref{fig:separation},
and the orbit of the particles is shown in the inset.
These results are similar to those in other CE simulations \citep[e.g.][]{Ohlmann+16a,Chamandy+19a,Chamandy+20}.

\subsection{Torque exerted on the particles by the gas}
\label{sec:torque}
In the top panel of Figure~\ref{fig:torque}, 
we show the evolution of the $z$-component of the torque applied on the particles about their CM (left axis, solid black), 
from $t\approx115\da$ onward.
The evolution of the orbital separation is shown for comparison (right axis, dashed).
The drastic reduction of the torque with time 
is consistent with previous studies involving different initial binary parameter values \citep{Chamandy+19b,Chamandy+20}.

In order to model this torque, 
we first need to determine which part of the gas contributes significantly to it.
This is done by integrating the torque out to various trial values of $\rho/\rho\ma$,
with $\rho\ma$ the maximum density in a slice through the orbital plane, for a given snapshot
(throughout the simulation, 
the density maximum occurs near the position of the AGB core particle). 
It is found that using the trial value $\rho/\rho\ma=0.006$ leads to a torque 
that deviates the least from the actual torque. 
In the top panel of Figure~\ref{fig:torque}, 
it can be seen that the torque exerted by gas inside the contour $\rho/\rho\ma=0.006$ (magenta) 
matches closely with that exerted by all the gas (black).
Henceforth, we refer to this as the threshold or critical density $\rho\crit=0.006\rho\ma(t)$.
Further justification for the value $0.006$ is provided in \ref{sec:rhocrit}.

However, it is important to note that $\rho\ma(t)$ depends on the choices involved in modifying the initial \textsc{mesa} profile 
and softening the gravity around the AGB core particle (Section~\ref{sec:methods}).
Therefore, the value of $\rho\ma$ (and by extension, the ratio $\rho\crit/\rho\ma$) has limited physical significance. 
This is not a major concern because the role of $\rho\ma$ is simply to provide a convenient normalization for the density.
We shall see below that the linear spatial scale of the contour $\rho=\rho\crit(t)$ is roughly proportional to 
the core-companion separation $a(t)$ and that the morphological evolution is approximately self-similar, 
as also seen in \citetalias{Escala+04}.
This self-similarity helps to explain
why setting $\rho\crit$ to be a fixed fraction of $\rho\ma$ turns out to be a fruitful approach.
A choice of reference density other than $\rho\ma$, 
e.g.,~that at the particle CM or averaged over the circle with radius $a/2$ 
centred on the particle CM,
could perhaps be used instead, 
and we plan to explore this possibility in future work.


\subsection{Fitting an ellipsoid to the threshold density contour}
\label{sec:fit}
To apply the uniform density ellipsoid model of Section~\ref{sec:spheroid},
it seems natural to use the ellipsoid that best approximates the contour at the threshold density $\rho=0.006\rho\ma$.
To simplify the fitting procedure,
we choose to fit two ellipses in perpendicular planes (which uniquely define an ellipsoid)
rather than directly fitting an ellipsoid.
We first fit the $\rho=\rho\crit$ contour in the orbital plane with a co-planar ellipse, 
treating the semi-major axis $A$, aspect ratio $B/A$ and phase shift $\Delta\phi$ as fit parameters.
We then fit the $\rho=\rho\crit$ contour in the plane perpendicular to the orbital plane
that contains the ellipse major axis (i.e.~the plane rotated by $\Delta\phi$ from the line joining the particles),
setting the length of the semi-major axis of this ellipse also equal to $A$
but fitting for the semi-minor axis $C$.
These two ellipses define an ellipsoid with major axis and one other axis in the orbital plane,
and the third axis perpendicular to the orbital plane.
The top two panels of Figure~\ref{fig:fitting} show an
example of such a fit, 
at the snapshot $t=188.7\da$, where the phase shift ($\Delta \phi$) is equal to $14.7^\circ$
(which also happens to be almost equal to the mean value of $\Delta\phi$ over the time domain analyzed,
as discussed below).
The blue ellipses show the fits to the $\rho=\rho\crit$ blue contour in the orbital plane (left) 
and orthogonal plane containing the ellipse major axis (right).

Our interest is in modeling the slow spiral-in phase, 
so the models cannot be applied right from $t=0$.
To determine the appropriate time domain to apply the ellipsoid model, 
we qualitatively judged when the density contours in the orbital plane resemble ellipses,
which led to the estimate $t>125\da$.
This choice is supported by quantitative analysis shown in Figure~\ref{fig:rmse_evol},
where we plot the root mean square error (RMSE) of our ellipsoidal fit, 
normalized to be between $0$ and $1$.
It is seen that the RMSE decreases below $0.1$ after $t=125\da$ and remains below this value thereafter.

In Figure~\ref{fig:params}~(left axis), 
we plot the evolution of key parameters of the fit:
the ratio of the semi-major axis to orbital separation $A/a$,
the ratio of semi-minor to semi-major axis $B/A$ ($C/A$ in the orthogonal plane),
and the phase angle $\Delta\phi$ (right axis).
The best fit values oscillate on the orbital period;
averaging over this variability we find that they are fairly constant, 
thought they also vary somewhat on longer timescales
for reasons not yet understood.
Furthermore, the values of $B$ and $C$ are comparable, 
which suggests that the ellipsoid can be approximated by a spheroid ($B=C$).
This is important because it allows us to make use of the simple expressions presented in Section~\ref{sec:spheroid}.
For the general case of ellipsoids, 
the quantities $\alpha$, $\beta$ and $\zeta$ have an integral form, which would complicate modeling.

\subsection{Mass evolution}\label{sec:mass}
In Figure~\ref{fig:mass} (left axis), 
we show the ratio of the mass of the gas enclosed by the $\rho=\rho\crit=0.006\rho\ma(t)$ 
surface to the mass $m=M\core+M_2$ of the particles.
This ratio decreases with time as the particles inspiral and is always $\ll1$.
This is the regime for which \citetalias{Escala+04} applied their uniform ellipsoid model.
The mean density inside the $\rho=\rho\crit$ surface is shown in blue (right axis). 
The mean density $\rhobar$ is roughly constant between $t=125\da$ and $t=270\da$,
but decreases sharply thereafter. 
This sudden decrease may be an artefact caused by the ratio $r\soft/a$ becoming too large,
which implies a larger fractional volume inside which the gravitational potential 
is underestimated (see also Section~\ref{sec:discussion_parameters}).

\subsection{To what extent can the uniform spheroid model reproduce the torque?}
\label{sec:model}
To model the torque, 
we calculate the $\alpha$ and $\beta$ parameters in equations (\ref{eq:torque_expression1}--\ref{eq:eccentricity})
separately for each frame, taking
\begin{equation}\label{AtildeBtilde}
  \Atilde = A, \qquad \Btilde = \frac{B+C}{2}.
\end{equation}
For the mean density $\rhobar$, 
we make the approximation that the torque applied by the heterogeneous ellipsoid defined by the surface $\rho=\rho\crit=0.006\rho\ma(t)$
can be replaced by that applied by a homogeneous spheroid with dimensions determined using equations~\eqref{AtildeBtilde}, 
and with density equal to the mean density $\rhobar$ inside the $\rho=\rho\crit$ contour.
The torque derived using equation~\eqref{eq:torque_z}, 
shown by the cyan line in the top panel of Figure~\ref{fig:torque}, 
reproduces quite closely the actual torque (black line in the same figure).

Further, by approximating $\Delta\phi$ and $B/A$ by their mean values over the whole time interval of analysis, 
$\langle\Delta\phi\rangle=14.9^\circ$ and $\langle \Btilde/\Atilde \rangle=0.654$, 
equation~\eqref{eq:torque_z} can be written as
\begin{equation}\label{eq:torque_z_mean}
    \tau_z \approx - \frac{1}{2} \pi G m \left(\betatilde-\alphatilde\right) 
    \cos \langle\Delta \phi\rangle \sin \langle\Delta \phi\rangle \rhobar a^2,
\end{equation}
where $\alphatilde=0.457$ and $\betatilde=0.771$ are the values of $\alpha$ and $\beta$ 
when $\Btilde/\Atilde$ is replaced by $\langle \Btilde/\Atilde\rangle$. 
Equation~\eqref{eq:torque_z_mean} was also used by \citetalias{Escala+04} to model equal-mass inspiralling BSMBHs 
(see Section~\ref{sec:intro}). 
We find that this approximation, shown as an orange line in the top panel of Figure~\ref{fig:torque},
aligns quite well with the actual torque (black), 
though it introduces a phase shift of about a half period.
The phase difference between the model and the actual torque
can be explained by the direct dependence of equation~\eqref{eq:torque_z_mean} on the square of the separation $a$,
which causes the modeled torque to be in phase with the separation.

\subsection{To what extent can solutions of the idealized binary perturbers
problem reproduce the torque?}
\label{sec:linear}
In this section, 
we compare the torque evolution in the simulation with that predicted by  the \citetalias{Kim+08} model, i.e.,
equation~\eqref{torque_Kim08}.
The \citetalias{Kim+08} model assumes that the unperturbed medium is uniform,
and in order to apply this model, 
one needs to specify the unperturbed density $\rho_0$ and unperturbed sound speed $c_\mathrm{s,0}$.
It is not clear how this should be done, so we tried different choices.
One possible choice is to set $\rho_0(t)$ to  
the value of the density on the surface that bounds the region that contributes significantly to the torque,
$\rho\crit=0.006\rho\ma(t)$, and $c_\mathrm{s,0}(t)$ to the mean sound speed interior to this surface.
In practice, we introduce an adjustable scale factor multiplying $\rho_0(t)$, which is expected to be of order unity,
so that we finally take
\begin{equation}
  \rho_0(t) = \xi\rho\crit(t), \qquad c_\mathrm{s,0}(t) = \overline{c}\sound(t),
\end{equation}
where $\xi=0.44$ and bar represents average inside the contour $\rho=\rho\crit(t)$.
Of the methods tried, this is the one that best reproduces the torque measured in the simulation.

The \citetalias{Kim+08} torque on the perturbers is then calculated using equation~\eqref{torque_Kim08_general_explicit},
and is shown by 
magenta lines in the bottom panel of Figure~\ref{fig:torque}, dashed for $\xi=1$ and solid for $\xi=0.44$.
The modeled torque using $\xi=1$ is seen to agree quite well with the simulation from $t=270\da$ onward, but this may be a coincidence since numerical effects may cause the torque to be overestimated in the simulation at late times (see Section~\ref{sec:discussion_parameters}).
For $t<250\da$, the \citetalias{Kim+08} torque obtained using $\xi=1$ is too large by a factor of order unity;
after multiplying the modeled torque by $\xi=0.44$, 
the model reproduces the simulation torque remarkably well, 
including (to varying degrees) the magnitude, amplitude and temporal phase evolution.

We also plot the particle Mach number $\Mach_\mathrm{p}$,
equal to the particle tangential speed in the particle CM frame 
divided by the volume-averaged mean sound speed $\overline{c}\sound$ 
averaged inside the contour $\rho=\rho\crit$,
in green using the right axis of the bottom panel of Figure~\ref{fig:torque}.
The value of $\Mach_\mathrm{p}$ can be seen to vary on the orbital period and also on longer timescales, 
but 
is restricted to the range $0.5$--$0.7$, so the particles' azimuthal motion is subsonic.

For completeness we now mention alternative choices for specifying $\rho_0(t)$ and $c_\mathrm{s,0}(t)$.
As a second approach, we tried setting $c_\mathrm{s,0}(t)$ to 
the value of $c\sound$ on the $\rho=\rho\crit$ surface, 
which leads to a larger Mach number and 
values of the torque that are about a factor of two larger than above.
A third approach we tried was to take both $\rho_0(t)$ and $c_\mathrm{s,0}(t)$ 
to be equal to the average values inside the $\rho=\rho\crit$ surface.
This produces results more similar to the 
chosen method (first approach) 
but the torque is similar in magnitude to what is obtained in the second approach.
From the bottom left panel of Fig.~\ref{fig:fitting}, 
one can see that the density at the contour $\rho=\rho\crit$ exceeds the background density.
Thus, to the extent that our simulation lines up with the \citetalias{Kim+08} model,
it makes sense that the best-fit value of $\rho_0$ for reproducing the torque is smaller than $\rho\crit$.

Fourth, we considered using the original profile of the AGB star by setting $\rho(t) = \rho(r)$ with $r=a(t)$,
and likewise for the sound speed.
This would be convenient if it worked, because then one would only need to know the initial conditions.
However, this choice produces torque values that are higher than those measured in the simulation by about an order of magnitude,
and a torque evolution that does not resemble the simulation even after dividing by a constant numerical factor.
The failure of this fourth approach suggests that it is not straightforward 
to model accurately the torque at late times using the initial stellar profile and \citetalias{Kim+08} model alone,
at least not for companions whose masses are appreciable compared to that of the giant's core.
This is perhaps not surprising because transfer of orbital energy and angular momentum 
to the envelope causes the profiles of density and sound speed to be drastically affected,
as shown in Figs.~\ref{fig:density} and \ref{fig:velocity} of \ref{sec:extra_sim_figs},
and, as we argue in Section~\ref{sec:discussion_assumptions}, 
the system loses memory of its previous state after about one sound-crossing time.
%

To summarize, 
the \citetalias{Kim+08} model reproduces the simulation remarkably well between $t\approx125\da$ and $t\approx250\da$ when
one adopts $\rho_0=0.44\rho\crit(t)$ and $c_\mathrm{s,0}=\overline{c}\sound(t)$, 
whereas at late times choosing $\rho_0=\rho\crit(t)$ produces closer agreement.
This is discussed further in Section~\ref{sec:discussion_parameters}.

\subsection{Comparison of the two idealized models}
\label{sec:correspondence}
As we have shown, 
the lagging spheroid model of \citetalias{Escala+04} and the double perturber model of \citetalias{Kim+08}
both reproduce the torque in the simulation reasonably well.
Thus, they would be expected to be consistent with one another.
Indeed, \citetalias{Kim+08} applied their model to the \citetalias{Escala+04} simulation,
arguing that their model can be used to explain the \citetalias{Escala+04} results if $\Mach\pert\sim0.6$.
The bottom-left panel of Figure~\ref{fig:fitting} is adapted from \citetalias{Kim+08} (their figure~1a).
The quantity plotted in colour is $\log(\mathcal{D})=\log[c_\mathrm{s,0}^2a/Gm)\lambda]$, 
which is equal to a constant plus the logarithm of the fractional density perturbation, 
$\lambda=(\rho-\rho_0)/\rho_0$ (with $\lambda\ll1$ assumed in \citetalias{Kim+08}), for $\Mach\pert=0.6$.
From the bottom panel of Figure~\ref{fig:torque} the relevant part of our simulation typically has $\Mach\pert\sim0.6$ as well.
To compare the ellipsoid model with that of \citetalias{Kim+08}, 
we annotate their figure by drawing an ellipse with parameters set equal to the time-averaged values from our simulation: 
$\langle A/a\rangle=1.47$, $\langle B/A\rangle=0.649$, and $\langle\Delta\phi=14.9^\circ\rangle$ (see also Figure~\ref{fig:fitting}).
The fairly close correspondence helps to explain why the \citetalias{Escala+04} and \citetalias{Kim+08} models produce similar values of the torque.
The time-averaged best fit ellipse in the orbital plane from our simulation 
is seen to align quite well with the structure obtained for $\log(\mathcal{D})$ by \citetalias{Kim+08}, 
though the latter is not precisely elliptical as the wakes form elongated tails.

The bottom-right panel of Figure~\ref{fig:fitting} shows $\log(\mathcal{D})$ from our simulation at $t=188.7$ d,
which is the same snapshot plotted in the top two panels.
To calculate $\mathcal{D}$, we assume $\rho_0=0.44\rho\crit=0.44\times0.006\rho\ma(t)$, 
as in the best-fit torque model discussed above.
The morphology is similar to that obtained by \citetalias{Kim+08}, 
but the wakes do not have tapered, elongated tails in the simulation.
We can also see that $\lambda$ is higher near the particles in the simulation than it is for \citetalias{Kim+08}.
This makes sense because \citetalias{Kim+08} assume the perturbers to be weak ($\lambda\ll 1$),
but in reality they are not (see Section~\ref{sec:discussion_assumptions} below).

\subsection{Spatial variation of key quantities and its evolution}
In Figure~\ref{fig:snapshots}, We show the time evolution of four different quantities, 
one per row. 
Columns show, from left to right, snapshots of slices through the orbital plane at $t= 138.9$, $208.3$, and $277.8\da$.
At these times, the particle separations are respectively $a=19.5$, $13.1$, and $9.4\Rsun$.

The top row shows the mass density in the orbital plane. 
The white contour is the equidensity surface ($\rho=\rho\crit=0.006\rho\ma$) that we fit with an ellipse, 
whereas the grey contours show a few other equidensity surfaces. 
The softening spheres of the AGB core (left) and companion (right) are represented by black circles. 
Note that the axis units are normalized to the current particle separation. 
Hence, the particle separation is always equal to $1$ in these units,
but the softening radius, which is constant ($r\soft=2.41\Rsun$), appears to grow with time in these normalized units.
The left and right snapshots are very similar despite $29$ orbits elapsing between them,
which implies a high degree of self-similarity in the problem as the separation decreases with time.
Note that the AGB core particle has a somewhat higher concentration of gas around it
than the companion particle, which 
is reasonable given that the AGB core particle starts the simulation
surrounded by dense gas, whereas the companion particle starts the simulation outside the AGB star.

The second row shows the 
spatially dependent Mach number of the 
particles,
equal to the 
particle tangential speed in the rest frame of the particle CM
divided by the local sound speed $c\sound$.
The gas in the torque-dominating region, delineated by the $\rho=\rho\crit$ contour, is subsonic.
Note the spiral shock, particularly evident in the middle panel,
where a sudden transition from blue (high sound speed) to red (low sound speed) 
is visible across the shock.
A jump in density across the shock can be seen in the middle panel of the top row 
where density contours almost overlap.

The third row shows the normalized difference between the angular velocity of the perturber and gas $(\Omega-\Omega\gas)/(\Omega+\Omega\gas)$,
where $\Omega=(\Omega_1+\Omega_2)/2$ is the mean angular speed 
of the particles and $\Omega\gas$ is the local angular speed of the gas.
Gas coloured red (blue) is rotating slower (faster) than the perturbers 
and gas coloured white is in corotation. 
Note that there is a fairly large region around each particle where the gas is corotating 
with the particles in their orbit.
This corotating gas exerts a torque on the particles because it preferentially lags them.
This assertion is supported by the fourth row, 
which shows the torque density (colour and contours). 
The black (white) contours show regions that contribute positively (negatively)
to the drag in the azimuthal direction.
We can see that while there is a large degree of symmetry, 
the positive contributions are slightly stronger (as evidenced by the slightly larger contours),
leading to a net drag, rather than a net thrust.

\section{SUCCESSES AND LIMITATIONS}
\label{sec:discussion}

We have seen that the simple models presented can be used to understand the torque evolution in our CE simulation.
The modeled torque and that measured from the simulation show a remarkable level of agreement.
However, for each model there are parameters -- e.g.~$\langle\Btilde/\Atilde\rangle$, 
$\rhobar(t)$, and $\langle\Delta\phi\rangle$ for the \citetalias{Escala+04} model 
and $\rho_0(t)$ and $c_\mathrm{s,0}(t)$ for the \citetalias{Kim+08} model -- that 
need to be estimated from the simulation,
and these estimates entail choices.
Further, both types of model rely on assumptions that drastically simplify the problem;
to what extent are these assumptions justified? 

\subsection{Ellipsoid model}\label{sec:discussion_ellipsoid}
The simplest version of the ellipsoid model is the constant aspect ratio, constant lag angle prolate spheroid model,
akin to the model used by \citetalias{Escala+04}.
This minimalistic model depends on three parameters, viz.~$\Atilde/a$, which is time-dependent, 
and $\Btilde/\Atilde$ and $\Delta\phi$, which are both treated as constants.
Recall that $\Atilde/a$ is relevant only because it affects the value of $\rhobar$ 
(see the discussion of Newton's third theorem in Section~\ref{sec:spheroid}).
This model does a reasonable job of producing the simulation results, particularly the period-averaged magnitude of the torque, 
which deviates from the torque measured in the simulation by $<16\%$ 
in the time domain studied (compare orange and black lines in the top panel of Figure~\ref{fig:torque}).
The model also does a good job of reproducing the amplitude and period of the torque variation,
though it does a poor job of reproducing the temporal phase.
This minimalistic version of the model is successful because the values of the parameters are fairly constant
(Figure~\ref{fig:params}).
However, it is not clear whether this is generally true, given the large diversity of common envelope parameter values in nature.
Moreover, we have not tried to motivate the values of the parameters from first principles and knowledge of the initial conditions
(but a place to start would be the \citetalias{Kim+08} model).

In any case, it is rather remarkable that the values of the parameters reported by \citetalias{Escala+04}, viz.~$\Btilde/\Atilde=1/2$ and $\Delta\phi=22.5^\circ$, 
are close to the time-averaged values we obtained, 
viz.~$\langle\Btilde/\Atilde\rangle=0.654$, and $\langle\Delta\phi\rangle=14.9^\circ$. 
In addition, we obtain $\langle \Atilde/a\rangle = 1.47$;
\citetalias{Escala+04} do not report this value, 
though they do find the behaviour to be self-similar, i.e.~$\Atilde/a\approx\const$.
In fact, when we measure the properties of the ellipse in their figure~8, we find $A/a\approx1.43$, 
$B/A=0.658$, and $\Delta\phi=16.1^\circ$, values that are remarkably similar (almost identical) to the ones we obtain.
This near-perfect correspondence is likely a coincidence, 
but it could suggest that the parameter values tend to converge to values that are fairly universal, 
in the sense that they only depend on a few basic quantities, such as $\Mach\pert$.
This possibility should be investigated in future work.

A correspondence of merger phenomena on vastly different scales 
is further suggested by the morphology in gas density seen in the merging galaxy simulations of \citet{Bortolas+18};
see the pre-merger snapshots shown in their figure~2, which are strikingly similar to those seen in our simulation.

For real CE events the ratio $q=M_2/M\core$ generally differs from unity, 
but the ellipsoid model cannot  be directly applied to the case of unequal perturber masses.  
For $q\ne1$, the density distribution is roughly pear-shaped \citep[e.g.][]{Gagnier+Pejcha25},
making it  more difficult to model the resulting gravitational potential.
However, the $q=1$ case provides the  opportunity 
to gain physical insight and interpret the simulation results with the simplest model.
Our result~--~that dynamical friction torque can be modeled as that due to a roughly constant-shape, 
$\sim a$-sized mass distribution
that lags the particle orbit by a $\sim$constant phase angle~--~is a principle that 
may carry over to unequal mass cases, albeit with a more complicated shape. 
A generalized approach analogous to our ellipsoid modeling would be  interesting to investigate
for $q\ne 1$ cases.
But the \citetalias{Kim+08} model and extensions thereof \citep{Kim10} 
are already more promising as a template
for modeling general cases since they already allow for $q\ne1$.

\subsection{Double perturber model}\label{sec:discussion_perturber}
Unlike the ellipsoid model which is partly phenomenological, 
the \citetalias{Kim+08} model is motivated from first principles but relies on various assumptions, 
discussed below in Section~\ref{sec:discussion_assumptions}.
Aside from these assumptions, 
an important limitation is that the parameters of the model, viz.~the density and sound speed of the unperturbed medium,
are difficult to constrain.

\subsubsection{Model parameters}\label{sec:discussion_parameters}
We found in Section~\ref{sec:linear} that using a density equal to that at the boundary of the torque-dominating region
and a sound speed equal to the average value inside this region leads to relatively good agreement with the torque in the simulation.
However, these choices are somewhat arbitrary.
Recall that the magnitude of the torque and amplitude of its variations for $155\da <t<250\da$ 
are reproduced remarkably well if one multiplies by the scaling factor $\xi=0.44$,
but after $t=270\da$, the agreement is good without needing to include a scaling factor.
It may be, however, that the factor of $0.44$ is needed in general, 
and that the torque in our simulation is overestimated for $t\gtrsim250\da$
because the ratio of separation to softening radius -- in the range $a/r\soft\sim3.5$--$4.5$ -- is too small at the end of the simulation. 
\citet{Gagnier+Pejcha25} find that the gravitational torque is overestimated
if $a\le3.6r\soft$ 
(where we have converted Plummer to spline softening radius by multiplying by $2.8$,
consistent with that work),
which is comparable to the values at the end of our simulation.
Moreover, \citet{Chamandy+19b} compared two identical global CE simulations similar to the one in this work
but differing in the softening radius by a factor of two, 
with the number of resolution elements per softening length fixed.
Consistent with the above suggestion, 
they found a slightly higher drag force for the model with larger softening radius, 
resulting in slightly smaller separation (their figure A1; see also figure~C2 and Table~C1 of \citealt{Chamandy+19a}). 
As mentioned in Section~\ref{sec:mass}, 
the sudden decrease in the mean density inside the surface $\rho=\rho\crit$ at $t\approx270\da$ 
may also be a consequence of the softening radius becoming too large with respect to $a$.

The torque is also affected by the resolution, and 
\citet{Gagnier+Pejcha25} 
find that one needs $\gtrsim47$
resolution elements per spline softening radius for convergence, 
whereas our simulation has a constant value of $r\soft/\delta_5\approx34$.
Less resolution per softening length tends to reduce the drag force 
\citep{Ohlmann16,Chamandy+19a,Gagnier+Pejcha25}.
Evidently, solutions can also depend on the locations, sizes, 
and shapes of the regions where the different levels of resolution are implemented 
(for example, one can compare models~D and F of \citealt{Chamandy+18}).
In this work, the resolution was chosen partly to ensure a reasonable level of energy conservation,
although by simulation end, the total energy had increased by $6\%$ (see \ref{sec:energy_conservation}).

In any case, 
the magnitude of the torque is reproduced if one 
sets $\rho_0=\xi\rho\crit$ in the \citetalias{Kim+08} model with $\xi=0.44$. 
While $\xi$ might approach $1$ at late times, this might only be because the torque in the simulation is overestimated.
Importantly, in addition to the overall magnitude,
the \textit{amplitude} and \textit{phase} of the torque are also reproduced remarkably well,
which is encouraging.
In summary, one obtains an excellent fit by taking $\Mach\pert$ as the value calculated using the mean sound speed
in the torque-dominating region and $\rho_0$ as $0.44$ times the value of the density at the boundary of this region.

\subsubsection{Consequences of neglecting various effects, as in the \citetalias{Kim+08} model}
\label{sec:discussion_assumptions}

The good agreement between the \citetalias{Kim+08} model and the simulation 
suggests that several approximations of the former are valid in the region of parameter space simulated.
For one, the \citetalias{Kim+08} model neglects gas self-gravity, 
which is justified in the regime considered since the point masses dominate the potential.
It is reasonable to neglect self-gravity if the ratio of the orbital period to the gas free-fall time,
$P\orb/t\ff\sim\sqrt{G\rhobar}/\Omega\sim\sqrt{\rhobar a^3/m}$ is small.
From Figure~\ref{fig:mass}, $\rhobar\lesssim 1.8\times10^{-5}\gcmcmcm$, 
and from Figure~\ref{fig:separation}, $a\lesssim23\Rsun$.
Thus, we obtain $P\orb/t\ff\lesssim0.2$ at $t\approx125\da$ and smaller thereafter, 
so self-gravity likely plays only a minor role.

Secondly, 
the \citetalias{Kim+08} model solves a linearized set of equations which depend on the assumption 
that the density perturbations are small, i.e.~the perturbers are weak.
This constraint is relaxed and the full nonlinear equations are solved by \citet{Kim+Kim09}, 
who study a single strong perturber in rectilinear motion,
and by \citet{Kim10}, who studies a single strong perturber in circular motion.
However, the double perturber case has not yet been studied in the nonlinear regime.
According to those papers, to be in the linear regime one requires $\mathcal{A}=GM\pert/c_\mathrm{s,0}^2r_\mathrm{s}\ll1$,
where $r_\mathrm{s}$ is the Plummer softening radius.
In our CE simulation, $\mathcal{A}\approx12$ (taking $r\soft=2.8r_\mathrm{s}$), so we are in the nonlinear regime.
However, for $\Mach\pert<1$, 
which is the regime in which we find ourselves (bottom panel of Figure~\ref{fig:torque}),
solutions of the drag force are found \textit{not to differ very much} from the linear case \citep{Kim+Kim09}.%
\footnote{For $\Mach\pert>1$, the torque in the nonlinear solutions is smaller than that in the linear solutions.}
This likely helps to explain why the \citetalias{Kim+08} model works so well to explain our CE simulation,
up to a scaling factor of order unity.
Nonlinear effects can also become important as $\mathcal{B}=Gm/c_\mathrm{s,0}^2a$ increases,
but for Mach numbers $>1$ \citep{Kim10}.
Even if the drag torque in the subsonic regime is well approximated by the linear equations, 
the morphology of the wakes produced by solving the full nonlinear equations, 
including weak shocks outside the orbit, 
matches better with what is obtained in our simulations (c.f.~the top two panels of figure~5 of \citealt{Kim10}).

A third idealization of \citetalias{Kim+08} is that the unperturbed density and sound speed are spatially constant,
which at first glance seems inconsistent with the CE case, 
where the density scale height of the original AGB profile 
can be comparable to the inter-particle separation during the simulation \citep{Chamandy+19b}.
However, the initial density profile gets drastically modified during the CE phase \citep[e.g.][]{Iaconi+17,Chamandy+20}, 
and hence seems fairly irrelevant for the late times explored in this work.
There seems to be little memory of the original profiles and flow pattern,
aside from their influence on how much gas is present near the perturbers at later times.

Another idealization of \citetalias{Kim+08} is that the CM of the binary is not in motion with respect to the envelope.
In our CE simulation, the particle CM moves with a speed of a few $\kms$ 
(as can be estimated from the inset of Figure~\ref{fig:separation}; see also \citealt{Chamandy+19a}).
As this speed is small compared to the orbital speed of $\sim60\kms$ (Figure~\ref{fig:velocity}), 
the CM motion is not expected to affect significantly the torque on the binary \citep{Sanchez-salcedo+Chametla14}.

In the \citetalias{Kim+08} model the semi-major axis of the binary orbit is fixed, 
whereas in our CE simulation it is decreasing with time. 
The timescale $\overline{a/\dot{a}}$, where bar here denotes average over an orbital period,
is much larger than the orbital period $P\orb$.
The flow is expected to reach a new steady state within a sound-crossing time \citep{Kim+Kim07,Desjacques+22},
\begin{equation}\label{sound_crossing}
  t\sound \sim \frac{a}{c\sound} \sim \frac{\Mach\pert a}{V\pert} \sim \frac{2\Mach\pert}{\Omega\pert} 
  = \frac{\Mach\pert P\orb}{\uppi} \lesssim 0.2P\orb,
\end{equation}
where the numerical estimate is made by taking $\Mach\pert\sim0.6$ (bottom panel of Figure~\ref{fig:torque}).
Thus, to a good approximation, the flow is able to continuously adjust to the local conditions and can be treated as quasi-steady.

In fact, the parameters ($\rho_0$ and $\Mach\pert$) also vary on the shorter timescale $P\orb$ due to the orbital eccentricity
\citep[e.g.][]{Szolgyen+22},
but relation~\eqref{sound_crossing} shows us that even this timescale is large compare to $t\sound$.
Hence, to a good approximation, 
one can treat the flow as quasi-steady and apply the \citetalias{Kim+08} model at any point in time
during the time interval studied.
This is only possible because the perturbers move subsonically at these times.
For recent studies of gaseous dynamical friction on perturbers in eccentric orbits 
see \citet{Buehler+24} and \citet{Oneill+24}.%
\footnote{Our results are consistent with the finding of \citet{Buehler+24} 
that for $e=0.3$ and $\Mach=0.6$, 
the analytical solution of \citet{Desjacques+22}, which assumes $e=0$, 
accurately reproduces the drag force in their simulations.}

\subsubsection{Limitations that are common to the simulation and \citetalias{Kim+08} model}
Thus far, we have discussed the limitations of the \citetalias{Kim+08} model for reproducing our simulation results,
but the \citetalias{Kim+08} model and our CE simulations also have certain assumptions \textit{in common} that are not fully realistic.
First, the equation of state is taken to be that of an adiabatic ideal gas with $\gamma=5/3$,
which means that ionization, recombination, radiative cooling, and radiative transfer are neglected.
Both hydrogen and helium are expected to be completely ionized near the perturbers,
the envelope is optically thick, 
and the orbital evolution is fairly insensitive to the equation of state \citep[e.g.][]{Sand+20,Chamandy+24}.
Thus, we do not expect that the torque evolution will depend very significantly on the equation of state and the inclusion of radiation
during the time interval considered.
However, at very late times not reached by our simulation the gas around the perturbers would become optically thin,
at which point these assumptions would be invalid and the modeled torque inaccurate.

Another assumption shared by \citetalias{Kim+08} and our simulation is that the perturbers do not accrete or launch jets/winds.
The accretion rate during CE evolution might, in some cases, 
attain hyper-critical values orders of magnitude larger than the Eddington rate,
in which case it could affect significantly the orbital separation through its effect on the gravitational torque \citep{Chamandy+18}.
However, 
this effect is unlikely to be important in cases where feedback prevents the accretion rate from greatly exceeding the Eddington rate.
Relevant work on dynamical friction drag on accreting perturbers in a gaseous medium is presented in
\citet{Ruffert96}, \citet{Sanchez-salcedo+Brandenburg01}, \citet{Gruzinov+20}, \citet{Prust+24}, and \citet{Suzuguchi+24}.
A somewhat related issue is the lack of sophistication in the modeling of the AGB core and companion, 
which are represented by point particles.
While there has been recent work on modeling the perturbers as spheres \citep{Prust20,Prust+Bildsten24}, 
this may not make a major difference since the mass distribution around the perturbers tends to be quasi-hydrostatic
in the CE simulations (e.g.~\citealt{Kim10}, \citealt{Chamandy+18}; see also the top row of Figure~\ref{fig:snapshots}),
mimicking, in a loose sense, the structure of a star.

Other effects that are not modeled are turbulence (possibly convective) and MHD effects.
Actually, turbulence seems to be captured to some extent in our CE simulations (\citealt{Chamandy+19b}; 
see also second row of Figure~\ref{fig:snapshots}, where patchy morphology suggests the presence of turbulence),
but turbulence is not considered in \citetalias{Kim+08}, 
though it may have a significant effect on the drag \citep{Lescaudron+23}.
Magnetohydrodynamics (MHD) is not considered in any of the above models, 
though magnetic fields can affect the dynamical friction force \citep{Sanchez-salcedo12,Shadmehri+Khajenabi12}.
While certain
CE simulation studies have found that magnetic fields only weakly affect the orbital evolution 
\citep{Ohlmann+16b,Ondratschek+22},
\citet{Vetter+25} find that the inspiral is significantly deeper at late times in their MHD run
as compared to their pure hydrodynamics run.
Moreover,
\citet{Ondratschek+22},
\citet{Vetter+24} and \citet{Vetter+25}
find that the magnetic fields are strong enough to produce powerful bipolar outflows.
More studies are needed to better understand the role of magnetic fields in CE evolution.

\subsection{Non-applicability to Very Early or Very Late Times}

The duration of our simulation is about $40$ orbits and the models are applied for the final $36$ orbits.
The symmetry that makes the \citetalias{Escala+04} and \citetalias{Kim+08} models applicable 
over the last roughly $36$ orbits is not present during the initial plunge-in.
At those times models based on wind-tunnel simulations 
\citep[e.g][]{Macleod+Ramirez-ruiz15a, Macleod+17,De+20}, 
or single perturbers \citep[e.g.][]{Kim+Kim07,Kim10} may be effective.
At very late times (perhaps $\gtrsim10^3$ orbits) the envelope will be largely ejected, 
and  self-similar evolution is unlikely to persist.
Some CE simulations  reveal an asymmetric, partly evacuated region in the orbital plane 
just outside the binary at these late times \citep{Gagnier+Pejcha25,Vetter+25}, 
so the radial density profile can be non-monotonic.
Different torque models may be required to explain these late stages of evolution.

\section{SUMMARY AND CONCLUSIONS}
\label{sec:conclusions}
We have presented a new global 3D CE simulation involving a $1.78\Msun$ AGB star and $0.53\Msun$ point-particle companion,
with the companion mass chosen to be equal to that of the AGB core particle.
This choice leads to a large degree of two-fold 
rotational symmetry in the orbital plane at late times
that facilitates interpretation.
We focus on understanding the gravitational (gas dynamical friction) torque on the particles, which causes them to inspiral,
and we limit our analysis to the slow-spiral in phase, i.e.~from approximately the fourth orbit ($125\da$) up until the $40$th orbit ($299\da$), when the simulation ends.
During this time, the gas inside the orbit is much less massive than the particles (Figure~\ref{fig:mass})
and the motion of the perturbers is subsonic (bottom panel of Figure~\ref{fig:torque} 
and second row of Figure~\ref{fig:snapshots}).
Our main results are as follows:
\begin{itemize}
  \item The torque is dominated by local gas, 
        out to about $1.5$ times the inter-particle separation from the particle CM 
        (compare black and magenta lines in the top panel of Figure~\ref{fig:torque} 
        and note the values of $A/a$ in Figure~\ref{fig:params}).
  \item We first interpreted our results using a model akin to that of \citetalias{Escala+04},
        modeling the torque on the particles as that produced by a uniform density ellipsoid
        rotating with the same angular velocity as the particles but lagging by a phase angle.
        The simplest version of this model invokes a prolate spheroid with constant axis ratio and lag angle.
        These models reproduce remarkably well the torque as a function of time,
        but contain parameters, viz.~the major axis to particle separation ratio (or, equivalently, the mean density), spheroid axis ratio, and lag angle,
        that needed to be estimated from the simulation.
  \item While the lag angle shows small variations on the timescale of the orbital period as well as on longer timescales,
        its relative constancy (Figure~\ref{fig:params}) shows that the density pattern of the gas approximately corotates with the binary orbit.
        Moreover, a significant component of the gas \textit{itself} is approximately in corotation with the orbit, 
        as can be seen in the second row of Figure~\ref{fig:snapshots},
        where the differential angular velocity of particles and gas is shown to be small.
  \item The best-fit lag angle, axis ratio, and semi-major axis to separation ratio are found to be remarkably close
        to those found by \citetalias{Escala+04} (their figure~8), who simulated BSMBHs; 
        it is unclear to what extent this points to universal behaviour in different astrophysics contexts or is more of a coincidence.
  \item \citetalias{Kim+08} developed a different, first principles model 
        to explain the torque on double perturber systems,
        and used it to explain the results of \citetalias{Escala+04}; 
        see the bottom row of Figure~\ref{fig:fitting} for an illustration of the consistency between the two types of model.
        We applied the \citetalias{Kim+08} model to our simulation and again found remarkable agreement with the measured torque
        magnitude, amplitude, and the temporal phase of the torque oscillations caused by orbital eccentricity.
        This level of agreement was obtained by setting the unperturbed density, a parameter of the model, 
        to $\xi=0.44$ times the density at the surface of the torque-dominating region around the particles,
        and by setting the perturber Mach number to the ratio of the particle tangential speed 
        to the mean sound speed in the torque-dominating region (bottom panel of Figure~\ref{fig:torque}).
        The best-fit value of $\xi$ 
        may be different for other common envelope systems with different initial conditions.
  \item 
        While our model reproduces the magnitude of the torque at very late times ($t>270\da$) with $\xi$ set to unity, 
        this is likely  a coincidence:  
        the torque at the very end of the simulation may be overestimated 
        because the separation to softening radius ratio has become quite small by then.
  \item Using instead the density and sound speed of the \textit{initial} AGB profile 
        to predict the unperturbed density and sound speed parameter in the \citetalias{Kim+08} model 
        does not result in good agreement.
        But perhaps a more complicated relationship between the initial profile and model parameters still exists, 
        and this possibility should be explored in future work.
  \item We argued in Section~\ref{sec:discussion_perturber} 
        that while there are many extra effects in the CE phase
        that are not included in the \citetalias{Kim+08} model, 
        e.g.~gas self-gravity, nonlinear density perturbations, eccentricity, and perturber CM motion,
        these effects affect the torque only weakly.
        Hence the high level of agreement observed between the \citetalias{Kim+08} idealized model and our CE simulation.
        The underlying reason for this may be that the flow adjusts on the sound-crossing timescale,
        which is shorter than the other timescales in the problem, e.g.~$t\sound<P\orb<t\ff<a/\dot{a}$.
  \item That nonlinear effects, in particular, 
        seem to be unimportant may be a consequence of the perturbers' motion being subsonic \citep{Kim+Kim09}.
  \item Moreover, we argued that other effects not included in any of the models studied in this work,
        such as accretion, ionization/recombination, radiation transport, and magnetic fields, 
        may have a fairly minor effect on the torque in most cases{, 
        although more study is warranted.}
\end{itemize}

Further work is needed to learn how to interpret the torque in other CE simulations
in terms of the \citetalias{Kim+08} model or future improvements thereof,
and constrain the model parameters. 
Our results strongly suggest that numerous simulations 
showing a long-term secular decrease in the binary separation were, in fact, correct.
They also hint at how one could eventually predict, perhaps with some accuracy, 
the timescale for this orbital tightening to occur.
Down the road, this may lead to an accurate assessment of CE outcomes 
(e.g., merger or envelope ejection) and, 
more generally, to increasingly accurate simple time-dependent models of CE evolution
\citep[e.g.,][]{Fragos+19,Bronner+24}.

\begin{acknowledgement}

We are grateful to the reviewer for a careful reading of the manuscript 
and helpful suggestions that improved the work.
We thank Andr\'{e}s Escala for words of encouragement and Shubhajit Saha and Abha Vishwakarma for discussions. This work used the computational and visualization resources in the Center for Integrated Research Computing (CIRC) at the University of Rochester and the computational resources of the Rosen Center for Advanced Computing (RCAC) at Purdue University, provided through allocations PHY230168 and TG-TRA210040 from the Advanced Cyberinfrastructure Coordination Ecosystem: Services \& Support (ACCESS) program \citep{Access}, which is supported by National Science Foundation grants \#2138259, \#2138286, \#2138307, \#2137603, and \#2138296, and the computational resources on Frontera of the Texas Advanced Computing Center (TACC) at The University of Texas at Austin, provided through TACC Frontera Pathways allocation AST23022. Frontera \citep{Frontera} is made possible by National Science Foundation award OAC-1818253. 

\end{acknowledgement}


\bibliography{refs}

\appendix

\section{Derivation of the torque on equal mass perturbers exerted by a corotating, 
lagging, uniform prolate ellipsoid}\label{sec:torque_derivation}

The general solution for the gravitational potential inside an ellipsoid of uniform density 
is given by equation~\ref{eq:potential}. 
Taking the angle between the major axis (of length $A$) of the ellipsoid and the line joining the binary components as $\Delta \phi$, 
we define the new coordinate system $(x', y', z')$ ($x'$ being the major axis of the ellipsoid) as follows:
\begin{equation}
\begin{aligned}
\begin{bmatrix}
x' \\
y' \\
z'
\end{bmatrix} &=
\begin{bmatrix}
\cos (\Delta \phi) & -\sin (\Delta \phi) & 0 \\
\sin (\Delta \phi) & \cos (\Delta \phi) & 0 \\
0 & 0 & 1
\end{bmatrix}
\begin{bmatrix}
x \\
y \\
z \\
\end{bmatrix} \\ 
&=
\begin{bmatrix}
x\cos (\Delta \phi) - y\sin (\Delta \phi) \\
x\sin (\Delta \phi) + y\cos (\Delta \phi) \\
z 
\end{bmatrix}
\end{aligned}
\end{equation}
The equation for the gravitational potential exerted by the gas on the particles $\Phi$ in the new coordinate system is
\begin{equation}
\begin{aligned}
    \Phi &= \pi G \rhobar \left[\alpha (x')^2 + \beta (y')^2 + \zeta (z')^2 - \chi \right] \\
    &= \pi G \rhobar 
    \left\{x^2\left[\alpha \cos^2(\Delta \phi) + \beta \sin^2 (\Delta \phi) \right]
    + xy\sin (2\Delta \phi) (\beta - \alpha)\right.\\
    & \quad \left.+ y^2\left[\alpha \sin^2(\Delta \phi) + \beta \cos^2(\Delta \phi)\right]
    + z^2 \zeta - \chi \right\}.
\end{aligned}
\end{equation}
The force on a particle of mass $M$ can be calculated by taking the negative gradient of the potential,
\begin{multline}
\begin{aligned}
    &\Vec{F} = - M \nabla \Phi
    = -\pi G \rhobar M \\
    &\times\bigg\{\left[2x\left(\alpha \cos^2(\Delta \phi) + \beta \sin^2 (\Delta \phi)\right) 
    + y\sin (2\Delta \phi) (\beta - \alpha) \right] \bfxhat  \\
    &+ \left[x\sin (2\Delta \phi) (\beta - \alpha) 
    + 2y\left(\alpha \sin^2(\Delta \phi) + \beta \cos^2(\Delta \phi)\right)\right]\bfyhat \\
    &+ 2\zeta z \,\bfzhat
    \bigg\}.
\end{aligned}
\end{multline}
The torque can then be calculated by taking the cross product of the position and force vectors. 
\begin{multline}
\begin{aligned}
    &\tau = \pi G \rhobar M \\
    &\times\bigg\{z\left[x\sin (2\Delta \phi) (\beta - \alpha) \right.\\
    &\left.+ 2y \left( \alpha \sin^2(\Delta \phi) + \beta \cos^2(\Delta \phi) -\zeta\right)\right] \bfxhat \\
    & -z\left[2x\left(\alpha \cos^2(\Delta \phi) + \beta \sin^2 (\Delta \phi) + \zeta\right)\right. \\
    &\left.+ y\sin (2\Delta \phi) (\beta - \alpha) \right] \bfyhat \\
    & + \left[\sin (2\Delta \phi) (\beta - \alpha)(y^2-x^2) \right.\\
    &\left.+ 2xy\Big(\alpha\cos(2\Delta\phi) - \beta\cos(2\Delta\phi)\Big) \right] \bfzhat\bigg\}
\end{aligned}
\end{multline}
When the origin of the coordinate system is taken to be the centre of mass of the particles 
and the particles are placed on the x-axis at $x= \pm a/2$ and $y=0$, 
the second term in the $z$-component of the torque vanishes 
and the spatial coordinates of the first term can be written in terms of the separation $a$. 
The $z$-component of the torque on each particle is thus given by 
\begin{equation}
    \tau_{z, 1/2} = - \frac{1}{4} \pi G \rhobar M \sin (2 \Delta \phi) (\beta - \alpha) a^2.
\end{equation}
Expanding the term $\sin(2\Delta \phi)$ into $2\sin(\Delta \phi)\cos(\Delta \phi)$ 
and writing the $z$-component of the total torque on both particles (total mass, $m = 2M$),
we obtain
\begin{equation}
    \tau_z = - \frac{1}{2} \pi G \rhobar m \sin (\Delta \phi) \cos (\Delta \phi) (\beta - \alpha) a^2,
\end{equation}
which is the same as equation~\eqref{eq:torque_z}.

\section{Determination of $\rho\crit$}\label{sec:rhocrit}

\begin{figure*}
    \centering
    \includegraphics[width=1\textwidth]{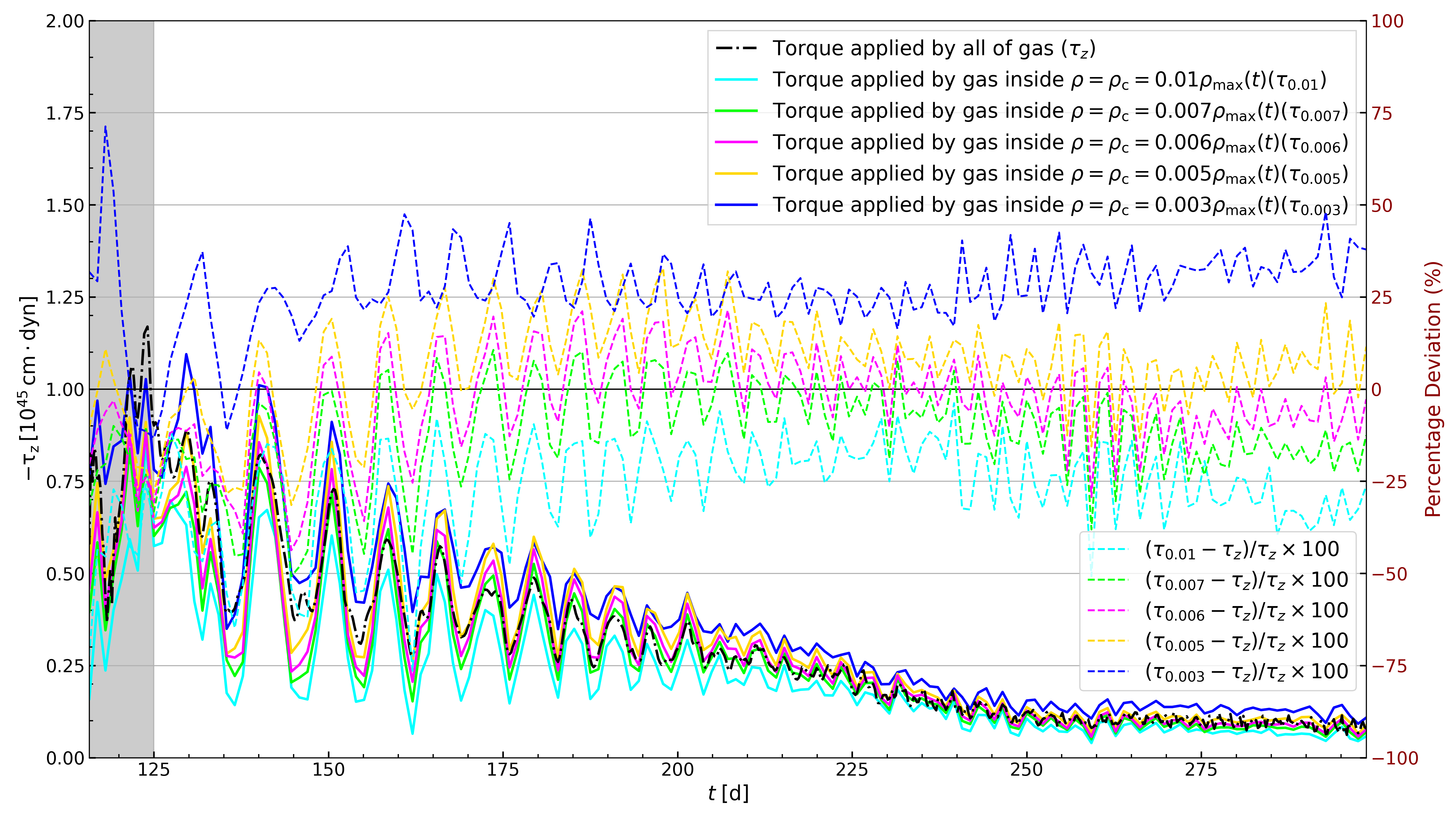}
    \caption{
    Comparison of the torque on the particles 
    contributed by gas inside the contour $\rho=\rho\crit= f\rho\ma$,
    with $\rho\ma$ the maximum density in the simulation,
    to the total torque contributed by all gas (dash-dotted black).
    The percent deviation from the total torque is shown on the right axis.
    For the torque models in this work we choose $f=0.006$,
    which gives the lowest time-averaged percent deviation of the values plotted for $t>125\da$.
    }
    \label{fig:rhocrit1}
\end{figure*}

The critical density $\rho\crit$ defines a geometric surface
that encloses the gas which approximately contributes all of the torque on the particles.
We also require this surface to be roughly ellipsoidal in shape
and small enough to be usefully applied to the torque models considered.
We model $\rho\crit$ as a fraction $f$ of the maximum density $\rho\ma$
(situated near the location of the AGB core particle for all snapshots),
and choose the value of $f$ which leads to a torque 
that is closest to the total torque exerted by all the gas in the simulation.
In Fig.~\ref{fig:rhocrit1}, we plot the total torque (black dash-dotted) 
and the torque for various values of $f$, as a function of time.
On the right axis, we plot the percent deviation.
The value $f=0.006$, shown in magenta, produces a torque closest to the total torque,
with an average percent absolute deviation of $10.6\%$, 
as compared to $13.9\%$ for $0.007$ and $12.3\%$ for $0.005$ in the range $t=125\da$ to $t=300\da$.
If we cap the range at $270\da$ to exclude the times when $r\soft/a$ is likely too large, 
$0.006$ still gives the smallest deviation ($11.1\%$).
Hence, we fix $f=0.006$ for this study.

\section{Evolution of gas density, sound speed, and particle velocities}\label{sec:extra_sim_figs}
\begin{figure}
    \centering
    \includegraphics[width=\linewidth]{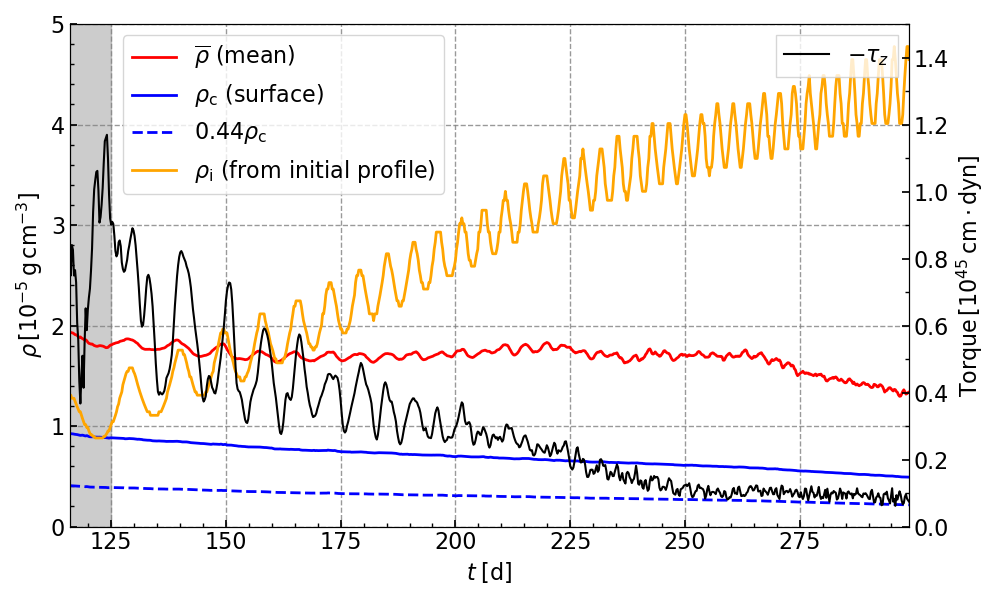}
    \caption{Mean density $\overline{\rho}$ inside the ellipsoid (red), 
    on the surface of the ellipsoid $\rho\crit$ (solid blue), 
    $\rho_0=0.44\rho\crit$ (dashed blue),
    and density from the initial profile at $a=r(t)$ (orange). 
    Total torque (black) is plotted on the right axis for reference.
    }
    \label{fig:density}
\end{figure}

In Figure~\ref{fig:density}, we plot the evolution of various gas densities in the simulation.
The mean density $\overline{\rho}$ inside the torque-dominating region bounded by the surface $\rho=\rho\crit=0.006\rho\ma$
is shown in red,
the surface density $\rho\crit$ in blue,
and the initial density profile of the AGB star $\rho(r)$ for $r=a(t)$ (where $a$ is the inter-particle separation) in orange.

\begin{figure*}
    \centering
    \includegraphics[width=\linewidth]{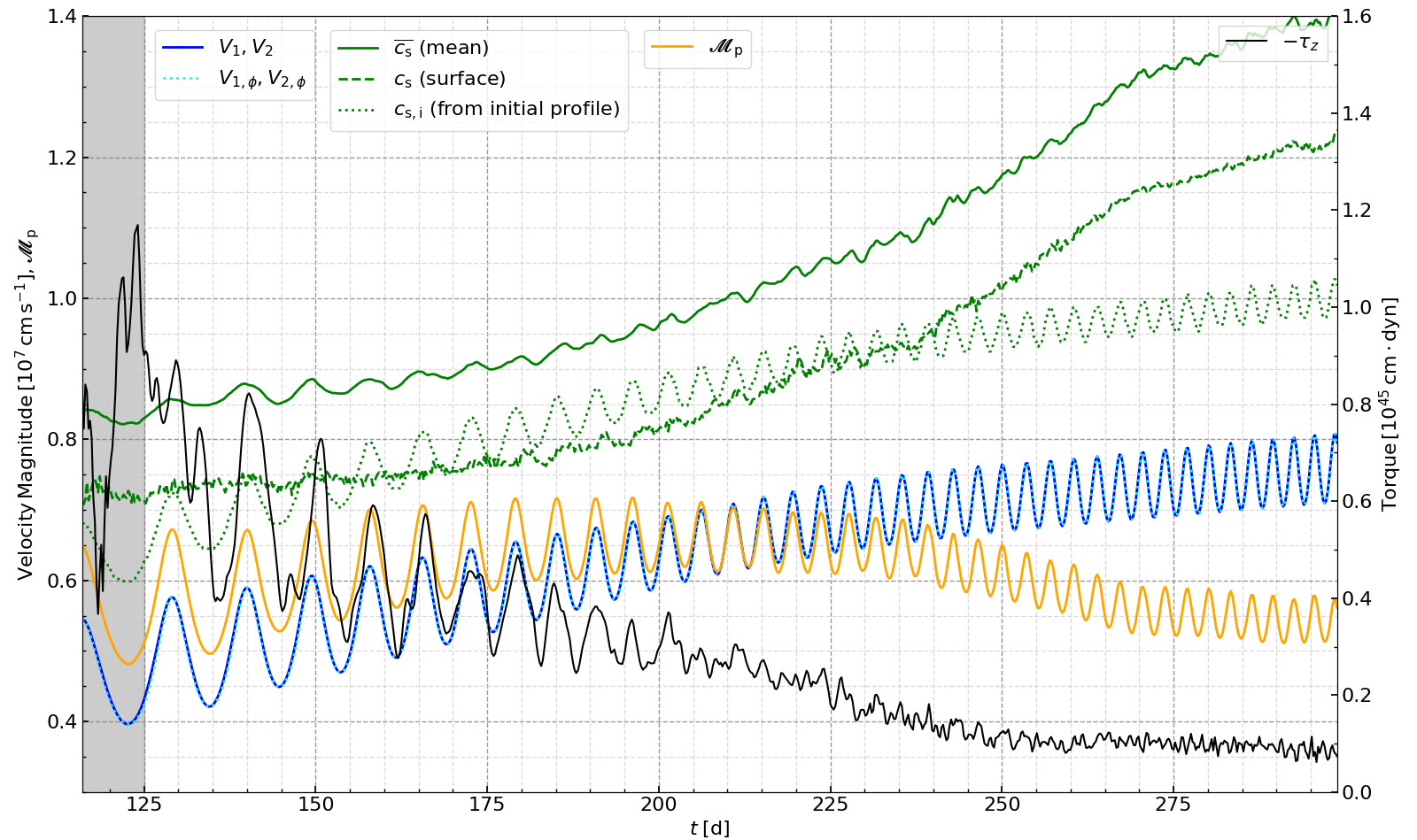}
    \caption{
    Speed of the particles in the particle CM frame (solid blue), 
    along with the 
    $\phi$-component of the particle velocity (dotted cyan)
    (the speeds of the particles are equal by definition in this frame).
    The magnitude of the $\phi$-component can be seen to coincide very closely with the total speed.
    The green lines show the adiabatic sound speed; 
    dashed, solid and dotted lines respectively show the mean value inside the surface $\rho=\rho\crit$, 
    the value on this surface, and the value at $r=a(t)$ in the original profile. 
    The solid orange line shows the 
    Mach number ($V_{\phi}/\overline{c}\sound$)  
    of the AGB core particle and companion, in the particle CM reference frame. 
    The total torque on the binary (black) is plotted on the right axis for reference.
    }
    \label{fig:velocity}
\end{figure*}

In Figure~\ref{fig:velocity}, we present the evolution of the magnitude and $\phi$-component of 
the velocity of each particle in the particle CM frame,
$V_1(=V_2)$ and $V_{1,\phi}(=V_{2,\phi})$,
are seen to be almost equal since the azimuthal orbital component dominates.
Also shown are the sound speed averaged inside the $\rho=\rho\crit$ contour, on the contour, 
or as obtained from the original profile by setting $r=a(t)$, where $a$ is the particle separation.

\section{Energy conservation}\label{sec:energy_conservation}
\begin{figure}
    \centering
    \includegraphics[width=\linewidth]{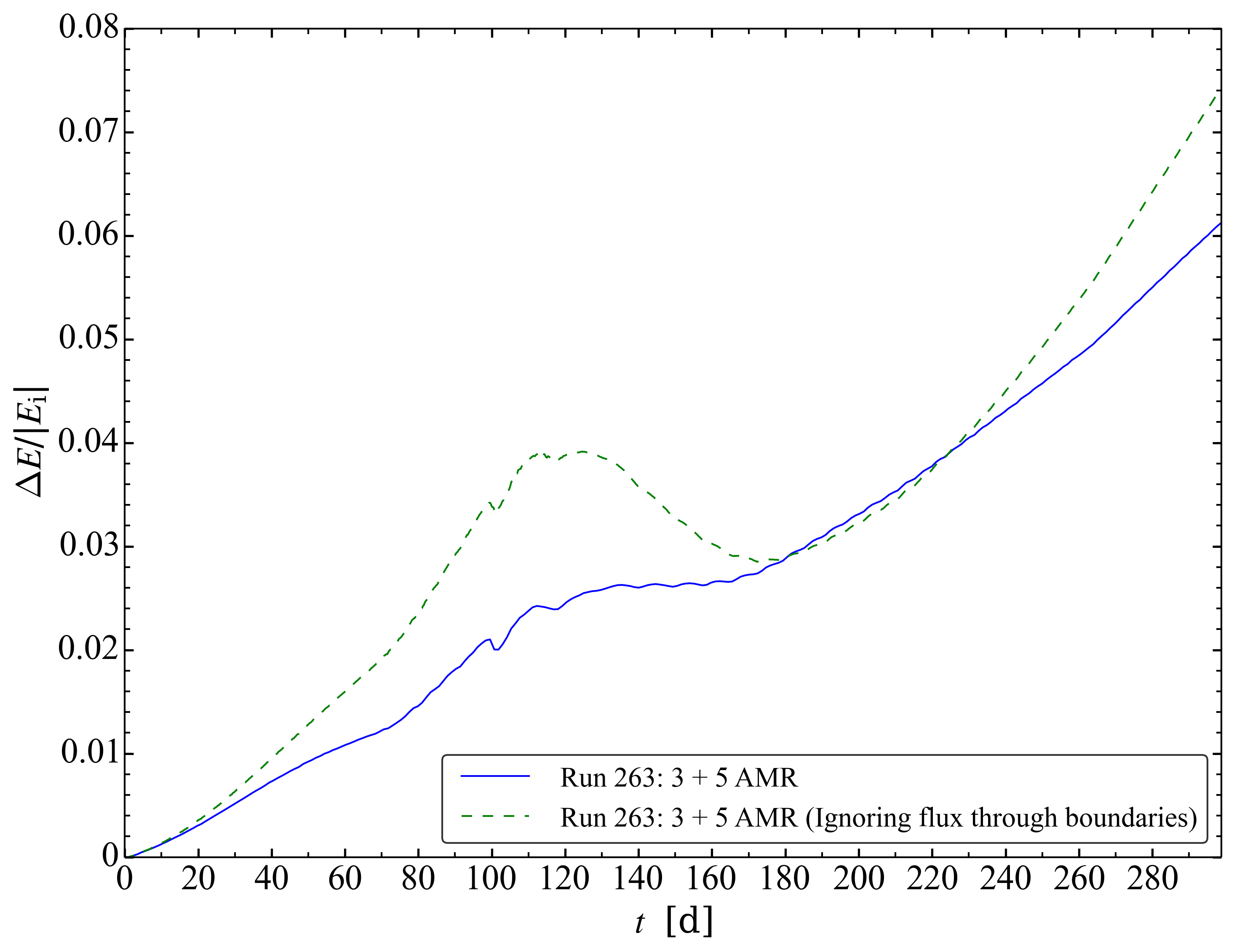}
    \caption{Percent change in energy is shown by the solid line. The dashed line considers the gas in the simulation domain only and does not account for the energy flux through the boundaries.}
    \label{fig:energy_percent_change}
\end{figure}

The percentage change in the total energy integrated over the simulation domain is presented in Figure~\ref{fig:energy_percent_change}.
The solid line takes into account the flux through the domain boundaries (and is therefore most relevant), 
whereas the dashed line does not.
We see that the total energy increases by about $6\%$ over the course of the simulation.

\section{Envelope unbinding}
\begin{figure}
    \centering
    \includegraphics[width=\linewidth]{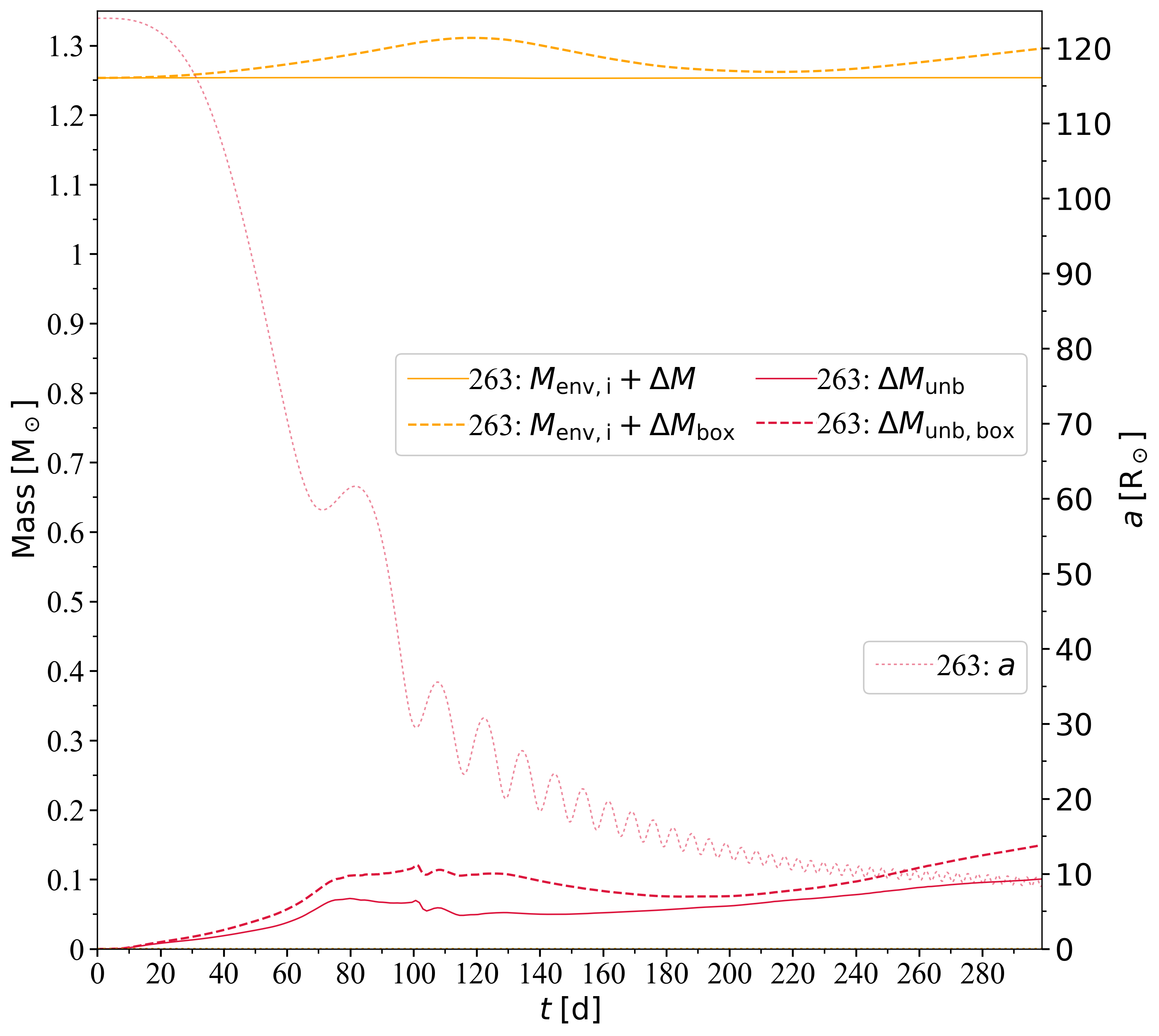}
    \caption{Unbound mass (solid red) and total mass (solid orange). 
    The dashed lines consider the gas in the simulation domain only and do not account for the mass flux through the boundaries.
    The dotted red line shows the particle separation, for reference (right axis).}
    \label{fig:unbound_mass}
\end{figure}

In Figure~\ref{fig:unbound_mass} we present the evolution of the unbound and total gas mass, 
integrated over the simulation domain.
Solid lines take into account the flux through the boundaries, but dashed lines do not.
One can see that total gas mass is conserved in the simulation (and the total mass, since the particle masses are kept constant).
For a comparison with a different simulation that is the same in all respects to the one studied in this work 
(including the same AGB primary star, initial separation, and initialization in a circular orbit) 
except that the companion is more massive ($0.98\Msun$ instead of $0.53\Msun$),
we refer the reader to figure~7 of \citet{Chamandy+20}.
The simulations have approximately the same duration but the $0.98\Msun$ companion simulation 
inspirals to a somewhat larger final separation and unbinds somewhat more mass than the $0.53\Msun$ simulation presented in this work.

\end{document}